\newif\ifabstract
\abstracttrue
 \abstractfalse 
\newif\iffull
\ifabstract \fullfalse \else \fulltrue \fi

\documentclass[11pt]{article}
\usepackage{amsfonts}
\usepackage{amssymb}
\usepackage{amstext}
\usepackage{amsmath}
\usepackage{xspace}
\usepackage{graphicx}
\usepackage{url}
\usepackage{graphics}
\usepackage{colordvi}
\usepackage{colordvi}
\usepackage{subfigure}

\advance \oddsidemargin by -0.8in
\newcommand{\myparskip}{3pt}
\parskip \myparskip

\usepackage[utf8]{inputenc}
\usepackage{mathrsfs}
\usepackage{amsmath,amsthm,bm}
\usepackage{xspace}
\usepackage{latexsym}
\usepackage{graphicx}
\usepackage{epsfig}
\usepackage{subfigure}
\usepackage{verbatim}
\usepackage{cases}
\usepackage{url}
\usepackage{amsfonts,latexsym}
\usepackage{setspace}
\usepackage{subfigure}
\usepackage{tikz}
\usepackage{pgfplots}
\usepackage{footnote}
\usepackage{enumitem}

\usepackage{algorithm}
\usepackage[]{algpseudocode}
\makeatletter
\newcommand{\algmargin}{\the\ALG@thistlm}
\makeatother
\newlength{\whilewidth}
\algdef{SE}[parWHILE]{parWhile}{EndparWhile}[1]
{\parbox[t]{\dimexpr\linewidth-\algmargin}{%
		\hangindent\whilewidth\strut\algorithmicwhile\ #1\ \algorithmicdo\strut}}{\algorithmicend\ \algorithmicwhile}%
\algnewcommand{\parState}[1]{\State%
	\parbox[t]{\dimexpr\linewidth-\algmargin}{\strut #1\strut}}

\makeatletter

\newcommand{\Rmnum}[1]{\expandafter\@slowromancap\romannumeral #1@}
\makeatother

\usepackage{array}
\usepackage{color}

\newtheorem{thm}{Theorem}[section]

\newtheorem{cor}[thm]{Corollary}

\newcommand{\cali}{\mathcal{I}}
\newcommand{\cala}{\mathcal{A}}

\DeclareMathOperator*{\argmax}{arg\,max}

\usepackage{algorithm,algpseudocode}
\usepackage{algorithmicx}
\usepackage{amsmath}
\usepackage{xspace}
\usepackage{color}
\usepackage{subfigure}
\usepackage{comment}
\usepackage{amsfonts}

\newtheorem{theorem}{Theorem}
\newtheorem{lemma}[theorem]{Lemma}
\newtheorem{assum}{Assumption}

\newcommand{\reg}{{\sf Regret}}

\newcommand{\otp}{\texttt{OTP}\xspace}	
\newcommand{\okp}{\texttt{OKP}\xspace}
\newcommand{\osc}{\texttt{OSC}\xspace}
\newcommand{\EAC}{\texttt{EAC}\xspace}
\newcommand{\okpalgalpha}{\texttt{OKP-Alg}(\alpha^{\texttt{on}})\xspace}
\newcommand{\okpalgalphaoffline}{\texttt{OKP-Alg}(\alpha^{\texttt{off}})\xspace}

\newcommand{\okpalg}{\texttt{OKP-Alg}\xspace}
\newcommand{\otpalg}{\texttt{OTP-Alg}\xspace}
\newcommand{\oscalg}{\texttt{OSC-Alg}\xspace}

\newcommand{\opt}{{\sf OPT}\xspace}
\newcommand{\alg}{{\sf ALG}\xspace}

\begin{document}

\title{Data-driven Competitive Algorithms for Online Knapsack and Set Cover}

\author{
Ali~Zeynali\thanks{University of Massachusetts Amherst. Email: {\tt azeynali@cs.umass.edu}.} \and
Bo~Sun\thanks{The Hong Kong University of Science and Technology. Email: {\tt bsunaa@connect.ust.hk}.}\and 
Mohammad~Hajiesmaili\thanks{University of Massachusetts Amherst. Email: {\tt hajiesmaili@cs.umass.edu}.} \and
Adam~Wierman\thanks{California Institute of Technology. Email: {\tt adamw@caltech.edu}.} \and
}

\begin{titlepage}
\maketitle

\thispagestyle{empty}

\begin{abstract}
The design of online algorithms has tended to focus on algorithms with worst-case guarantees, e.g., bounds on the competitive ratio.  However, it is well-known that such algorithms are often overly pessimistic, performing sub-optimally on non-worst-case inputs.  In this paper, we develop an approach for data-driven design of online algorithms that maintain near-optimal worst-case guarantees while also performing learning in order to perform well for typical inputs. Our approach is to identify policy classes that admit global worst-case guarantees, and then perform learning using historical data within the policy classes. We demonstrate the approach in the context of two classical problems, online knapsack and online set cover, proving competitive bounds for rich policy classes in each case. Additionally, we illustrate the practical implications via a case study on electric vehicle charging.
\end{abstract}

\end{titlepage}

\section{Introduction}
\label{sec:intro}
 
 
As the adoption of machine learning (ML) in infrastructure and safety-critical domains grows, it becomes crucially important to ensure ML-driven algorithms can provide guarantees on their performance. Competitive analysis of online algorithms has been a remarkably successful framework for developing simple algorithms with worst-case guarantees. The ultimate goal in this framework is to devise algorithms with the best possible \textit{competitive ratio}, which is defined as the worst-case ratio between the cost of an online algorithm and that of the offline optimal.  However, because the focus is on worst-case guarantees, the algorithms developed are typically conservative and do not learn in a data-driven manner.  The result is that worst-case optimized algorithms tend to under perform for typical scenarios where worst-case inputs are uncommon.  This phenomenon is wide-spread, but the importance of provable guarantees for safety and robustness means relaxing worst-case assumptions is undesirable in many settings.  An important open question is how to achieve the ``best-of-both-worlds'', both near-optimal performance in typical settings, which requires data-driven adaptation, and a near-optimal competitive ratio.

Toward this goal, there have been substantial efforts to improve the performance of competitive algorithms using predictions~\cite{chen2016using,chen2015online,antoniadis2020online},  ML advice~\cite{lykouris2018competitive,angelopoulos2019online,purohit2018improving,rohatgi2020near}, and advice from multiple experts~\cite{gollapudi2019online}.
In these approaches the goal is to allow online algorithms to use (potentially noisy) predictions (or advice) about future inputs.  
Such predictions capture the fact that, often, something is known about the future that could be used to improve the performance in typical cases. 
These approaches have been successfully applied to design competitive algorithms that perform near-optimally in the typical case in settings such as the ski-rental problem~\cite{purohit2018improving,kodialam2019optimal,angelopoulos2019online}, online optimization with switching costs~\cite{chen2016using}, online caching~\cite{lykouris2018competitive,rohatgi2020near}, and metrical task systems~\cite{antoniadis2020online}, to name a few.
However, in this literature, the power of ML models has been leveraged to \textit{first}, predict the future input, and \textit{then} modify the algorithms to use this additional input to further improve the performance. In this way, the learning and algorithmic parts are decoupled, and practical improvements could be obtained only when fine-grained predictions of individual inputs are (nearly) perfect.

The idea of this paper is inspired by the fact that practitioners typically prefer to learn from the coarse-grained patterns observed in previous problem instances and then optimize over a class of algorithms that achieves high performance given the coarse-grained patterns. This approach has been of interest of empirical studies for a long time~\cite{semke1998automatic,leyton2009empirical,kotthoff2012evaluation,winstein2013tcp,akhtar2018oboe,de2019erudite}. However, developing a theoretical understanding of this approach has received attention only recently, after a seminal work of Gupta and Roughgarden~\cite{gupta2017pac,gupta2020data} and its follow-ups~\cite{kleinberg2019procrastinating,balcan2018dispersion,alabi2019learning,cohen2017online}.


Inspired by the above high-level idea, in this paper, we focus on developing \textit{data-driven competitive algorithms} with worst-case guarantees. The prior literature on online algorithms with ML advice combines online algorithms with learning in order to learn the uncertain input. This work, in contrast, designs policy classes of algorithms that ensure competitive guarantees for all algorithms in the policy class and then learns the best algorithm based on historical data directly. The result is an adaptive algorithm, tuned based on historic data, that ensures worst-case robustness. Further, it allows the richness of the policy class to be balanced with the performance in the typical case -- if the policy class is broadened then the competitive ratio grows but there is a potential to learn a better algorithm for the common case. 


Realizing the potential of this approach requires two steps.  First, developing a policy class of online algorithms whose worst-case competitive ratios are bounded.  This is in contrast to the typical style of analysis in online algorithms that seeks an individual algorithm with an optimal competitive ratio. The form of the policy class is crucial, since it should be broad enough to allow adaptation in application settings, but still provide near-optimal competitive guarantees for all policies in the class; thus balancing worst-case guarantees with performance in the typical case.  Second, the approach requires developing ML tools that can learn and adapt from historical data to select a policy from the policy class.  This second task is standard, and can be approached with a variety of tools depending on the setting (as discussed in Section~\ref{sec:online_learning}).  

More formally, our goal in this paper is to derive policy classes of online algorithms such that all policies in the class have bounded degradation in terms of worst-case competitive ratio. To that end, we introduce the \emph{degradation factor} $\texttt{DF}(A(\theta))$ of an algorithm as the worst-case performance degradation of an algorithm $A(\theta)$, parameterized by parameter $\theta$, with respect to a baseline algorithm:
\begin{equation}
    \label{eq:competitive_ratio_pol}
    \texttt{CR}(A(\theta)) \leq \texttt{DF}(A(\theta))\texttt{CR}(A(\theta_0)),
\end{equation}
where $A(\theta_0)$ is the baseline algorithm that could be the one that achieves the optimal competitive ratio, the algorithm in literature with the best known competitive ratio, or simply the current algorithm used in practice.
Then, to characterize the degradation factor of a policy class, we define a class of $\phi$-\textit{degraded} algorithms as one where every algorithm in the class has degradation factor that is no larger than $\phi$, i.e., ${\mathcal{P}(\phi) = \{\theta |\texttt{DF}(A(\theta)) \leq \phi\}}$. Given such a class, in runtime, ML tools can be used to learn algorithm $A(\hat\theta)$, where $\hat\theta$ is the learned parameter, to optimize the performance while being ensured of a worst-case loss of at most $\phi$. 

As compared to the framework introduced in~\cite{gupta2017pac}, where the data-driven algorithm design is proposed as a general framework of improving an algorithm, this work, to the best of our knowledge, is the first that brings the idea of learning the best online algorithm among a policy class of algorithms that all have provable a worst-case competitive ratio guarantees. In addition, the degradation factor introduced in the paper is a novel performance metrics that can explicitly characterize the worst-case performance loss of online algorithms due to the data-driven algorithms selection. 



\paragraph{Summary of contributions.} To introduce and develop the above framework, we first consider, as a warm-up example, the classic ski-rental problem (Section~\ref{sec:ski}).  We provide a simple, concrete construction of a policy class of algorithms for this problem. Then, we characterize the degradation factor as well as $\phi$-degraded policies within the policy class. This result enables a trade-off between optimizing the typical case while ensuring near-optimal worst-case performance. 

After this warm-up, our main technical results focus on illustrating the approach in two important online problems: online knapsack (\okp) (Section~\ref{sec:okp}) and online set cover (\osc) (Section~\ref{sec:osc}).  For both \okp and \osc, we develop policy classes, explicitly characterize the degradation factor achieved by all policies in the class. For \osc, using a synthetic input, we provide intuition  about how the proposed framework can leverage the coarse-grained structure of inputs to learn the best policy. This is a new design space that could not be captured by online algorithms with ML advice that only utilize the fine-grained prediction of future inputs. 

Deriving the above results requires addressing two sets of technical challenges: (i) determining how to \textit{choose the right policy class} of algorithms, which requires a delicate balance between the worst-case guarantees and the learning design space for practical improvement, and (ii) determining how to \textit{bound the competitive ratio of a class of algorithms in a parametric manner}, which is not always a straightforward extension of the analysis of the classical analysis.
For example, in our analysis of the policy class for \okp, we have to identify and analyze two worst-case instances that differ from the classical analysis of \okp.

Then, in Section~\ref{sec:online_learning}, inspired by ideas in~\cite{balcan2018dispersion}, we discuss how to cast a regret minimization online learning problem for the selection of the best algorithm given a policy class of online algorithms. This section shows how to efficiently learn the best policy from the class, and highlights our work as a complementary view in the emerging space of data-driven algorithms.


Finally, to demonstrate the potential for achieving both data-driven adaptation and worst-case guarantees in practical settings, we apply our approach to the application of online admission control of electric vehicles in a charging station, which is an extended version of \okp. Our experiments consider two months of electric vehicle data-traces from a parking lot at Caltech~\cite{LeeLiLow2019a}, and show that the approach can improve the observed performance by 13.7\%, with only 3\% of instances having worse performance than the worst-case optimized online algorithm.

\section{Warm-up: The Ski-Rental Problem}
\label{sec:ski}
To illustrate our approach using a simple example, we start with the classic ski-rental problem~\cite{karlin1986competitive,borodin2005online}, in which a skier goes skiing for an unknown number of days. On each day, the skier can either rent skis at a unit price or buy one at a higher integer price of $p > 1$, and ski for free from then on. The uncertainty is the number of days the skier will ski.  Our focus is on deterministic algorithms, for simplicity, and the best known deterministic algorithm uses a \textit{break-even point} and rents the first $p - 1$ days before buying on the $p$th day.  It is straightforward to see that this algorithm is 2-competitive, which is optimal.  However, such a choice of the break-even point is overly conservative in typical situations. 

Consider now a policy class of algorithms $A(b)$ with $b\in \mathcal{B} = \{1, 2, \dots\}$ defines the policy class. The parameter $b$ is the number of renting days and can be optimized based on historical data in order to improve the typical performance, at the cost of an increased competitive ratio. The following theorem characterizes the degradation factor of $A(b)$ followed by a corollary characterizing the policy class of $\phi$-degraded algorithms.


\begin{thm}
\label{thm:cr_ski}
Let $A(b), b \in \mathcal{B} = \{1, 2, \dots\}$ be a policy class of algorithms. The degradation factor of algorithm $A(b)$, with respect to the baseline $A(p)$, the optimal 2-competitive algorithm, is ${\emph{\texttt{DF}}(A(b)) = 1/2 + \max \{p/2b,b/2p\}}$.
\end{thm}
Note that with $b=p$, $\texttt{DF}(A(b)) = 1$, and $A(b)$ reduces to the algorithm that optimizes the competitive ratio. A proof follows quickly from standard results and is given in \S\ref{app:ski} in the supplementary material. Immediately from Theorem \ref{thm:cr_ski}, the $\phi$-degraded policy class is characterized as follows. 
\begin{cor}
\label{theo:ski_rent}
Let $\phi\geq 1$. The policy class of $\phi$-degraded algorithms for the ski-rental problem is determined by $\mathcal{B}(\phi)= [p/(2\phi-1),p(2\phi-1)]$.  That is, the degradation factor of any $A(b)$ with $b\in\mathcal{B}(\phi)$ is no larger than $\phi$: $\emph{\texttt{DF}}(A(b))\leq \phi, b\in\mathcal{B}(\phi)$  .

\end{cor}

\begin{figure}
    \centering
    \includegraphics[scale=0.5]{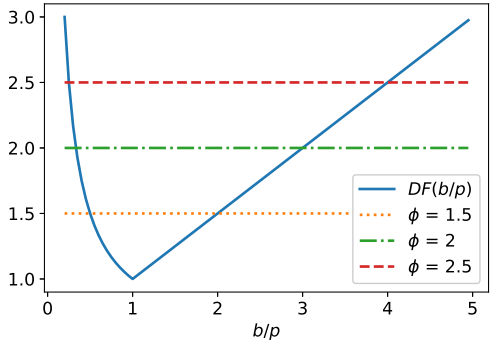}
    \vspace{-4mm}
    \caption{The degradation factor as a function of normalized parameter $b/p$ for the ski-rental problem}
    \label{fig:PoL_ski}
\end{figure}

 To elaborate on typical design guidelines that follow the above analysis, in Figure~\ref{fig:PoL_ski}, we plot the degradation factor as a function of the normalized parameter $b/p$. With $b/p>1$, the competitive ratio degrades gracefully, while with $b/p<1$ the competitive ratio degrades drastically. The figure also highlights a few $\phi$-degraded policy classes. For example, the 2-degraded policy class, which leads to a  degradation factor of at most 2, is $\mathcal{B}(2) = p\times[1/3,3]$. This wide range shows that data-driven learning of a policy can be effective at optimizing the typical case. We discuss how to perform the data-driven learning in Section \ref{sec:online_learning}.



\section{The Online Knapsack Problem}
\label{sec:okp}
The goal of the online knapsack problem (\okp) is to pack items that are arriving online into a knapsack with unit capacity such that the aggregate value of admitted items is maximized. In each round, item $i \in [n] = \{1,\dots, n\}$, with value $v_i$ and weight $w_i$, arrives, and an online algorithm must decide whether to admit or reject $i$ with the objective of maximizing the total value of selected items while respecting the capacity.
Given items' values and weights $\{v_i,w_i\}_{i\in[n]}$ in offline,  \okp can be stated as
\begin{subequations}
\label{eq:OKP_form}
    \begin{eqnarray*}
      {[\textrm{Offline}\ \okp]} \quad \max&  \sum\nolimits_{i\in[n]} v_i x_i, \\
      \textrm{s.t.,}& \sum\nolimits_{i\in[n]} w_{i}x_i\leq 1, \\ 
      \textrm{vars.,} & \quad x_i \in \{0,1\}, \quad i \in [n],
    \end{eqnarray*}
\end{subequations}
where the binary variable $x_i =1$ denotes the admission of item $i$ and $x_i=0$ represents a decline.  
In an online setting, the admission decision $x_i$ for item $i$ must be made only based on causal information, i.e., the items' values and weights up to now $\{v_j,w_j\}_{j\in[i]}$ and previous decisions $\{x_j\}_{j\in[i-1]}$. 
Since there exists no online algorithm with a bounded competitive ratio for the general form of \okp \cite{zhou2008budget}, we focus on \okp under the following two standard assumptions, e.g., \cite{zhang2017optimalposted,zhou2008budget}. 

\begin{assum}\label{assum:infitesimal}
The weight of each individual item is much smaller than the unit capacity of the knapsack, i.e., $w_i \ll 1, \forall i\in[n].$
\end{assum}
\begin{assum}\label{assum:bounded-value-density}
The value-to-weight ratio (or value density) of each item is lower and upper bounded between $L$ and $U$, i.e., $L \leq v_i/w_i \leq U, \forall i\in[n].$
\end{assum}

Assumption \ref{assum:infitesimal} naturally holds in large-scale systems, including the case study of this work in the context of electric vehicle charging. Assumption \ref{assum:bounded-value-density} is to eliminate the potential for rare items that have extremely high or low-value densities. 
This version of \okp has been used in numerous applications including online cloud resource allocation~\cite{amarante2013using,zhang2017online}, budget constrained bidding in keyword auction~\cite{zhou2008budget}, and online routing~\cite{buchbinder2009design}. Also, as rigorously stated in appendix, \okp is closely related to the one-way trading problem (\otp)~\cite{el2001optimal} in the sense that the optimal competitive ratios of \okp and \otp are the same and both can be achieved via threshold-based algorithms. Consequently, our  results also hold for \otp. 


Under Assumptions \ref{assum:infitesimal} and \ref{assum:bounded-value-density}, and using $\gamma := U/L$ to represent the fluctuation of the value density, the optimal competitive ratio of \okp is $\ln(\gamma)+1$, in the sense that no (deterministic or randomized) online algorithms can achieve a smaller competitive ratio~\cite{zhou2008budget}.
It has also been shown that a threshold-based algorithm can achieve this optimal competitive ratio~\cite{zhou2008budget,zhang2017optimalposted}. 
Let $\Psi(z):[0,1]\to[L,U]\cup \{+\infty\}$ denote a threshold function. 
The online algorithm for \okp admits an item $i$ only if its value density is no less than the threshold value at current utilization level $z_{i-1}$, i.e., $v_i/w_i \ge \Psi(z_{i-1})$.
With the threshold function designed in~\cite{zhang2017optimalposted,zhou2008budget}, this algorithm is proved to be $\left(\ln(\gamma)+1\right)$-competitive. 

\subsection{A Competitive Policy Class for \okp}
\label{sec:fine-tune}

\begin{figure}
\begin{minipage}[b]{0.45\textwidth}
    \centering\includegraphics[width=0.99\linewidth]{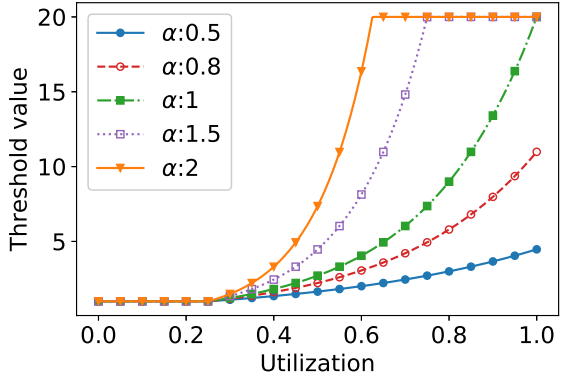}
    \vspace{-3mm}
    \caption{Threshold function $\Psi_\alpha$ in Eq.~\eqref{eq:thresholdf_prime} with different values of $\alpha$ for \okp; $\gamma = 20$.}
    \label{fig:threshold-function}
\end{minipage}\hspace{1mm}
\begin{minipage}[b]{0.45\textwidth}
    \centering\includegraphics[width=0.93\linewidth]{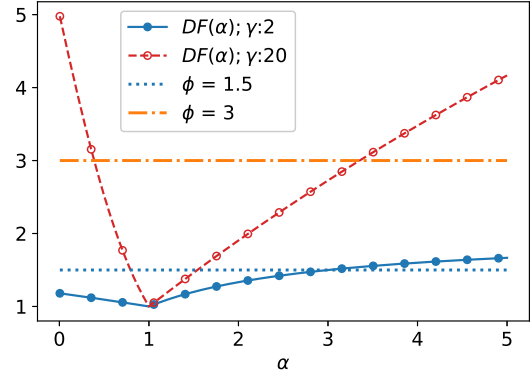}
    \caption{Degradation factor {\texttt{DF}$\left(\okpalg(\alpha)\right)$} for different value of $\gamma$ for \okp.}
    \label{fig:price-of-learning}
\end{minipage}\hspace{2mm}
\end{figure}

We consider a policy class of algorithms $\okpalg(\alpha)$, $\alpha \in \mathcal{A} : = \{\alpha|\alpha > 0\}$. $\okpalg(\alpha)$ is a threshold-based algorithm whose threshold function $\Psi_{\alpha}$ is parameterized by $\alpha$ as follows
\begin{equation}
\label{eq:thresholdf_prime}
  \Psi_{\alpha}(z) =
    \begin{cases}
      L & z \in [0, T), \\
      \min \left\{ U, L e^{\alpha(z/T - 1)}\right\}& z \in [T,1),\\
      +\infty & z = 1,
    \end{cases} 
\end{equation}
where $T = 1/(\ln(\gamma)+1)$ is a utilization threshold where all items will be admitted.
Figure~\ref{fig:threshold-function} depicts $\Psi_\alpha$ with respect to multiple parameter $\alpha$.
We can interpret $\Psi_{\alpha}(z)$ as the marginal cost of packing items into the knapsack when its utilization is $z$. 
This threshold function can then be used to estimate the cost of using a portion of the knapsack capacity and $\okpalg(\alpha)$ aims to balance the value from the item and cost of using the capacity. 
Particularly, upon the arrival of item $i$, $\okpalg(\alpha)$ makes the admission decision by solving a pseudo-utility maximization problem
\begin{equation}\label{p:okp}
x_i^* = \argmax_{x_i\in\{0,1\}} v_i x_i - \Psi_\alpha(z_{i-1})w_i x_i,
\end{equation}
where $z_{i-1} = \sum_{j\in[i-1]}w_jx_j^*$, is the cumulative capacity utilization of the previous $i-1$ items. Given Assumption \ref{assum:infitesimal}, $\Psi_\alpha(z_{i-1}) w_i$ estimates the cost of packing item $i$. The online algorithm admits an item only if its value density is high enough such that a non-negative pseudo-utility can be obtained. The $\okpalg(\alpha)$ is summarized as Algorithm~\ref{alg:ota}.
Note that the parametric algorithm $\okpalg(\alpha)$ is a generalization of the classic threshold-based algorithm~\cite{zhang2017optimalposted,zhou2008budget}.
When $\alpha = 1$, $\Psi_\alpha$ recovers the threshold function designed in~\cite{zhang2017optimalposted} and $\okpalg(1)$ can achieve the optimal competitive ratio. 

To construct a competitive policy class for \okp, one could consider various ways of parametrizing the threshold $\Psi_\alpha$. For instance, one could parameterize the minimum threshold $T$ of the flat segment (${\alpha\in[0,T)}$) and/or the increasing rate of the exponential segment ($\alpha\in[T,1)$) to increase aggressiveness of $\okpalg(\alpha)$. Our results focus on the exponential rate, which provides a richness for tuning without incurring significant degradation in the competitive ratio.
Our first result characterizes the degradation factor of $\okpalg(\alpha)$ with respect to the worst-case optimized algorithm \okpalg($1$).

\begin{algorithm}[!t]
	\caption{$\okpalg(\alpha)$: A parametric algorithm for \okp}
	\label{alg:ota}
	\begin{algorithmic}[1]
	    \State \textbf{Input:} threshold function $\Psi_\alpha$, initial utilization $z_0 = 0$;
		\While{item $i$ arrives}
		\State determine $x_i^*$ by solving the problem \eqref{p:okp} given the current utilization $z_{i-1}$;
        \State update utilization $z_i = z_{i-1} + w_ix_i^*$ 
		\EndWhile
	\end{algorithmic}
\end{algorithm}

\begin{thm}
\label{thm:DF_OKP}
The degradation factor of \emph{$\okpalg(\alpha)$}, $\alpha\in\mathcal{A}$, with respective to \emph{$\okpalg(1)$} is
\begin{align}
    \emph{\texttt{DF}$\left(\okpalg(\alpha)\right)$} = 
    \begin{cases}
    \frac{\alpha \gamma}{\alpha +  \gamma - 1}& \alpha \in[1,+\infty),\\
    \frac{\alpha \gamma }{\alpha + \gamma ^{\alpha} - 1} & \alpha \in (0,1).
    \end{cases}
\end{align}
\end{thm}
Figure~\ref{fig:price-of-learning} illustrates the degradation factor as the parameter $\alpha$ varies. In \S\ref{app:okp} of the supplementary document, we provide insights on the growth of the degradation factor and the rigorous proof of Theorem~\ref{thm:DF_OKP}. The proof leverages the fact that, in the worst-case scenario, the arriving items can be divided into two batches. $\okpalg(\alpha)$ only admits the first batch of items while the offline optimal solution only admits the second batch. Depending on the parameter $\alpha$, there exist two cases: when $\alpha \in [1,+\infty)$, the value densities of the first batch of items are exactly equal to the marginal cost of packing each item upon their arrivals and the total weight of the admitted items is $z^u$, at which the maximum value density is reached, i.e., $\Psi_\alpha(z^u) = U$. Then the second batch of items with value density $U-\epsilon$ ($\epsilon > 0$) arrives and their total weight is $1$. The competitive ratio of this case can be derived as $(U-\epsilon)/\int_{0}^{z^u}\Psi_\alpha(s)ds$.
When $\alpha \in (0,1)$, since the maximum marginal cost of packing items is less than the maximum value density, i.e., $\Psi_\alpha(1) < U$, the first batch of items will occupy the whole capacity of the knapsack. Thus, in this case, the competitive ratio becomes $U/\int_{0}^{1}\Psi_\alpha(s)ds$. Substituting $\Psi_\alpha$ gives the competitive ratios in the two cases.
In order to adaptively tune $\alpha$ with worst-case performance guarantees, data-driven models can be proactively built to ensure a degradation factor no larger than $\phi$ by restricting $\alpha$ in a $\phi$-degraded policy class. The following corollary characterizes the $\phi$-degraded policy class of $\okpalg(\alpha)$.

\begin{cor}
Let $\phi \in [1,\gamma)$. The policy class of $\phi$-degraded algorithms for \okp is
\begin{align}
\mathcal{A}(\phi) = \left[-\frac{\phi}{\gamma - \phi}- \frac{W_{\phi,\gamma}}{\ln(\gamma)} , \frac{\phi(\gamma - 1)}{\gamma - \phi}\right],
\end{align}
where $W_{\phi,\gamma} := W\left(-\frac{\ln(\gamma) \phi}{\gamma - \phi} e^{-\ln(\gamma)\phi/(\gamma -\phi)}\right)$ and $W(\cdot)$ is the Lambert function.
\end{cor}
The $\phi$-degraded policy class models a tradeoff between the performance in typical settings and the robustness in worst-case scenarios. A larger $\phi$ corresponds to a weaker worst-case guarantee but provides a larger space for potentially learning an algorithm that can work better in common cases. 

\section{The Online Set Cover Problem}
\label{sec:osc}
The classical (unweighted) version of online set-cover problem (\osc) is defined as follows. Let $\mathcal{I}$, with $n := |\mathcal{I}|$, be the ground set of elements, each indexed by $i$.  Let $\mathcal{S}, m := |\mathcal{S}|$, be a family of sets, such that each set  $s \in \mathcal{S}$ includes a subset of elements in $\mathcal{I}$. 
In the online problem, a subset of elements $\mathcal{I}' \subseteq \mathcal{I}$ arrives element-by-element over time. Once element $i$ arrives, the algorithm must cover $i$ by selecting one (or more) sets containing $i$. The online algorithm knows $\mathcal{I}$ and $\mathcal{S}$ in advance, but not $\mathcal{I}'$.
The ultimate goal is to pick the minimum possible sets from $\mathcal{S}$ in order to cover all elements in $\mathcal{I}'$. Given $\mathcal{I}'$, \osc can be formulated as follows.
\begin{subequations}
\label{eq:OSC_form}
    \begin{eqnarray*}
      {[\textrm{Offline}\ \osc]} \ \min & \sum\nolimits_{s\in \mathcal{S}} x_s, \\
      \textrm{s.t.,} & \ \sum\nolimits_{s|i \in s} x_s \geq 1, \ i\in \mathcal{I}', \\ 
      \textrm{vars.}, & \ x_s \in \{0,1\}, \ s \in \mathcal{S},
    \end{eqnarray*}
\end{subequations}
where the binary variable $x_s=1$ if set $s$ is chosen, and $x_s=0$, otherwise. The offline \osc problem is NP-hard and the online version is even more challenging due to the uncertainty of elements to be covered. The \osc problem was first introduced in~\cite{alon2003online} and has been proved to be the core problem in numerous real-world applications such as online resource allocation~\cite{pu2018online,wang2017online}, crowd-sourcing~\cite{sheng2012energy,bagaria2013optimally}, and scheduling~\cite{pananjady2015online}, etc.
In the literature, there exists several online algorithms for \osc~\cite{alon2003online,buchbinder2009online}. Our design is based on~\cite{alon2003online} where the proposed algorithm is based on a specifically-designed potential function. Upon arrival of an element, the algorithm adds some subsets to cover the current element, and meanwhile ensures the potential function is non-increasing. This algorithm achieves a competitive ratio of $4\log n (2+\log m)$. We primarily focus on developing a class of algorithms for \osc by extending the algorithm in~\cite{alon2003online}, and characterize the degradation factor with respect to it. Also, we provide  insights about how to learn an online algorithm from the class.


\subsection{A Competitive Policy Class for \osc}
In this section, we introduce $\oscalg(\theta)$, a parametric policy class for \osc that generalizes the existing algorithm, from~\cite{alon2003online}.  
The core of the policy class is a parameter $\theta$ that determines the rate at which the subsets should be covered. The algorithm works as follows. Let $w_s$ represent the weight of set $s\in\mathcal{S}$.  These weights evolve during the execution of $\oscalg$. Further, let $w_i = \sum_{s \in \mathcal{S}_i} w_s$ be the weight of element $i$, where $\mathcal{S}_i$ is the set of all subsets containing $i$, and define $\mathcal{I}^{\texttt{sel}}$ and $\mathcal{S}^{\texttt{sel}}$ as the running sets of covered elements and chosen subsets by the algorithm during its execution. The algorithm maintains a potential function $\Phi$ for the uncovered elements defined as $\Phi = \sum_{i \notin \mathcal{I}^{\texttt{sel}}} n^{2w_i}.$


The details on how the algorithm proceeds are summarized in Algorithm~\ref{alg:oa}. Briefly, once a new element arrives, if the element is already covered, the algorithm does nothing. However, if the new element is uncovered, the algorithm first updates the weight of the set containing $i$ according to Lines~\ref{algline:mink} and~\ref{algline:update}, and then selects at most $2\theta \log(n)$ subsets from $\mathcal{S}_i$ such that the potential function does not increase. Note that, with $\theta=2$, the algorithm degenerates to the existing algorithm in~\cite{alon2003online}, which covers at most $4\log n$ subsets in each round. For intuition behind potential functions and updating the weights, we refer to~\cite{alon2003online}. 
The following theorem characterizes the degradation factor of $\oscalg(\theta)$ using~\cite{alon2003online} as the baseline.

\begin{algorithm}[!t]
	\caption{$\oscalg(\theta)$: A parametric algorithm for \osc}
	\label{alg:oa}
	\begin{algorithmic}[1]
		\State \textbf{initialization:} potential function $\Phi$, $\theta > 1$, and $w_s = \frac{1}{\theta m }, s\in\mathcal{S}$, $\mathcal{I}^{\texttt{sel}} = \emptyset$,  $\mathcal{S}^{\texttt{sel}} = \emptyset$
		\While{ element $i$ arrives}
		\If{$w_i \geq 1$, i.e., $i$ is in  $\mathcal{I}^{\texttt{sel}}$}
        \State do nothing
        \Else { do the weight-augmentation:}
        \State find the minimum k such that $\theta^k \times w_i > 1$ \label{algline:mink}
        \State update $w_s \leftarrow \theta^k \times w_s, s \in \mathcal{S}_i$ \label{algline:update}
        \State select at most $2\theta \log(n)$ subsets from $\mathcal{S}_i$ such that value of potential function  does  not  increase  if we add these subsets to $\mathcal{S}^{\texttt{sel}}$. Add $i$ to  $\mathcal{I}^{\texttt{sel}}$ \label{algline:select}
        \EndIf
		\EndWhile
	\end{algorithmic}
\end{algorithm}


\begin{thm} 
\label{thm:df_osc}
The degradation factor of \emph{$\oscalg(\theta)$} with respect to \emph{$\oscalg(2)$} is 
\begin{align}
\label{eq:df_osc}
    \emph{\texttt{DF}}(\emph{\oscalg}(\theta)) = \theta \bigg[ \frac{2\log \theta + \log m}{2\log \theta (2+\log m)} \bigg].\end{align}
\end{thm}
As expected, this theorem recovers the result of ~\cite{alon2003online} for the case of $\theta=2$, obtaining the competitive ratio of $\texttt{CR} = (4 \log n)(\log m + 2)$ which is $O(\log n \log m)$. 
The proof (in \S\ref{app:osc} of supplementary) is based on finding a feasible region for $\theta$ such that the number of iterations in the weight augmentation is upper bounded, and ensuring the feasibility of the operation in Line~\ref{algline:update} of \oscalg. Also, note that with a slight change in the setting and assuming that the frequency of elements in subsets is bounded by $d$, \oscalg can be straightforwardly extended to the case with a competitive ratio of $(2\theta \log n)(2 + \log d/\log \theta)$. 

\begin{figure}
\begin{minipage}[b]{0.45\textwidth}
    \centering\includegraphics[width=0.95\linewidth]{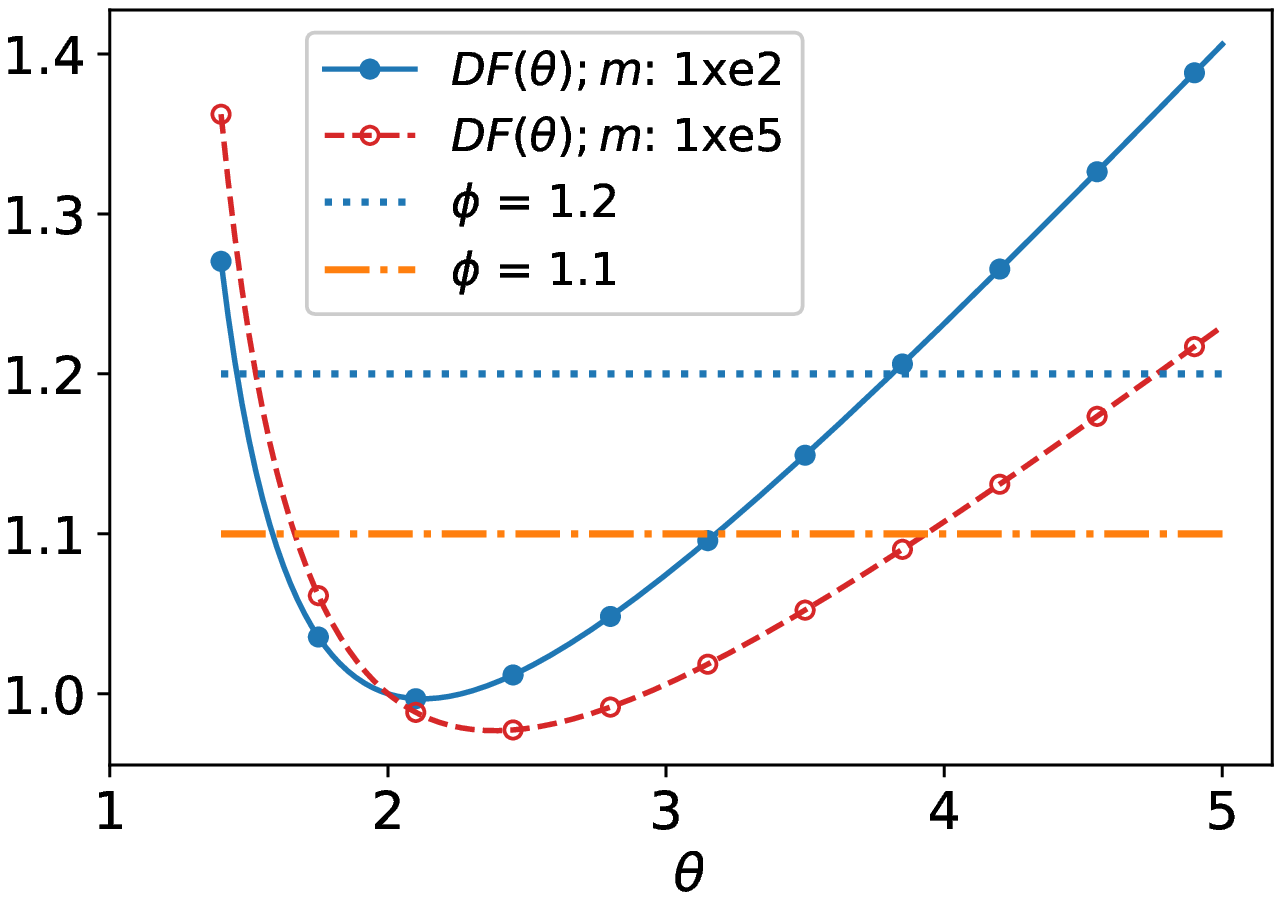}
    \vspace{-1mm}
    \caption{Degradation factor {\texttt{DF}$\left(\oscalg(\theta)\right)$} with different values of $m$ for \osc}
    \label{fig:osc_beta_theta}
\end{minipage}\hspace{1mm}
\begin{minipage}[b]{0.45\textwidth}
    \centering\includegraphics[width=0.95\linewidth]{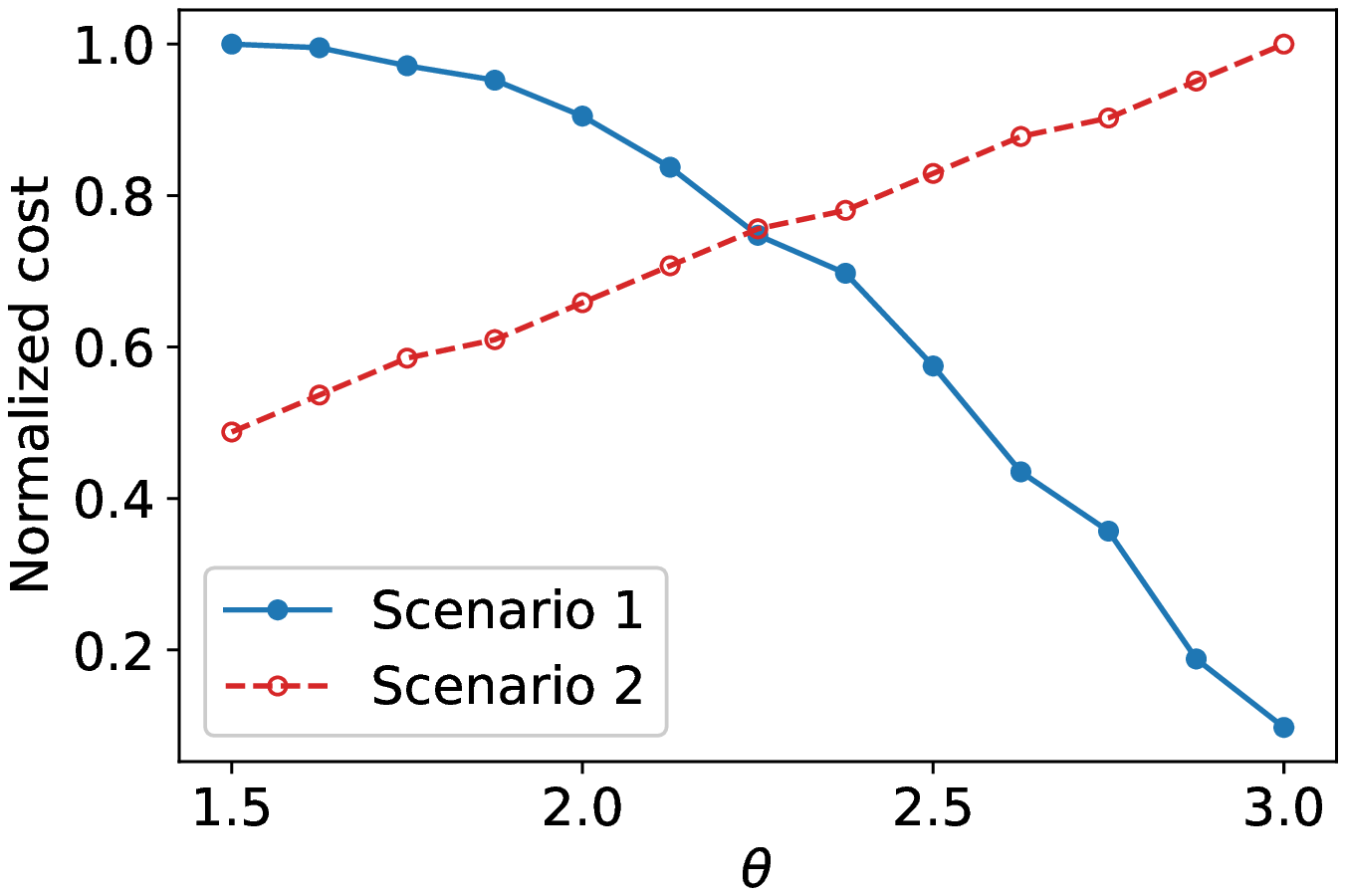}
    \vspace{-1mm}
    \caption{Normalized cost of $\oscalg(\theta)$ for two different problem scenarios}
    \label{fig:osc_intuition}
\end{minipage}\hspace{2mm}

\end{figure}

The next step is to characterize the $\phi$-degraded policy class of algorithms, which involves calculations of the inverse of Eq.~\eqref{eq:df_osc}, and can be expressed by the $r$-Lambert function, i.e., the inverse of $f(x)=x e^x + rx$. However, for clarity, in Figure~\ref{fig:osc_beta_theta}, we plot the degradation factor for two different values of $m$. Also, this figure shows 1.2 and 1.1-degraded policy classes. For example, it shows that with $m=10^5$, by tuning $\theta = [1.67,3.93]$, the competitive ratio will be degraded by at most 10\%. 

Last, to further highlight the practical difference of our proposed framework with the existing prediction-based online algorithms, we provide intuition on how the coarse-grained structure of input provides insights on finding the best policy. 
A key algorithmic nugget of \oscalg is on the number of subsets selected in Line~\ref{algline:select}. The higher the value of $\theta$, the higher the probability of selecting more subsets in runtime.  However, depending on the overlap of elements in subsets, a higher value $\theta$ might be good or bad. Specifically, with higher $\theta$, i.e., selecting more subsets in each round, is beneficial if the overlap of appearing elements in subsets is high since the algorithm covers some \emph{useful} subsets in advance. We refer to this as ``Scenario 1''. As ``Scenario 2'', consider the case in which the overlap between elements in subsets is small. In this scenario, smaller $\theta$ might be better since in each round the algorithm will cover only subsets that are needed for the current element. To illustrate the impact of learning the right value of $\theta$ under Scenarios 1 and 2, in Figure \ref{fig:osc_intuition}, we report an experiment with 20 random instances of \osc with $n = 120$, $m = 3200$, and 80 elements appeared on the input. The result illustrates our intuition above, i.e., larger $\theta$ favors Scenario 1 and smaller $\theta$ favors Scenario 2. Our construction of policy class is able to capture these scenarios for learning the best policy. Instead, the existing prediction-based algorithms can only use fine-grained ML advice that predicts the upcoming elements in near future. 

 \section{Data-driven Algorithm Selection as Online Learning}
\label{sec:online_learning}


The previous sections have constructed competitive policy classes of online algorithms for ski-rental, \okp, and \osc.  Given these classes, the next question is how to adaptively select the parametric algorithms using a data-driven approach. This task falls into the emerging framework of data-driven algorithm design, which has been introduced in~\cite{gupta2017pac,gupta2020data} and followed by~\cite{balcan2018dispersion} by framing it as a principled online learning problem.  In this section, we will discuss how the algorithm selection problem within our policy classes can be formulated as an online learning problem.

Consider the problem of adaptively selecting the parametric algorithm $A(\alpha)$ from a $\phi$-degraded policy class $\cala(\phi)$ in a total of $M$ rounds, where each round $t$ corresponds to an instance of the underlying online problem. 
At the beginning of round $t\in[M]$, we choose a parameter $\alpha_t \in \cala(\phi)$ and run $A(\alpha_t)$ to execute the instance of this round $\cali_t$.
Let $R_t(\cali_t,\alpha_t)$ denote the total reward in round $t$. 
Suppose the instances of all rounds are known from the start, the best fixed parameter
$\alpha^{\rm off}_c$ is given by $\alpha^{\rm off}_c = \arg\max_{\alpha\in\cala(\phi)}\sum_{t\in[M]} R_t(\cali_t,\alpha)$. 
The problem of choosing the parameter is to design an online learning algorithm that can determine $\alpha_t$ in an online manner and minimize the regret of the adaptively selected parameter $\alpha_t$ with respective to the best fixed parameter $\alpha^{\rm off}_c$, i.e., 
\begin{align*}
    \reg_M(\alpha^{\rm off}_c) = \sum_{t\in[M]}R_t(\cali_t,\alpha^{\rm off}_c) -  \sum_{t\in[M]}R_t(\cali_t,\alpha_t).
\end{align*}

To perform online learning for algorithm selection, we can apply results from prior literature.  One approach is to convert the infinite set of parameters $\cala(\phi)$ into a finite set $\tilde{\cala}(\phi)$ using discretization, numerically evaluate the reward function, and then apply existing online learning algorithms to determine the parameter selection.  Depending on the computational complexity of evaluating the reward function, we may choose ${\sf Hedge}$ algorithms \cite{freund1997decision} for full information feedback, i.e., known $R_t(\cali_t,\alpha), \forall \alpha \in \tilde{\cala}(\phi)$, or ${\sf EXP3}$ algorithms \cite{auer2002nonstochastic} for bandit information feedback, i.e., when the reward $R_t(\cali_t,\alpha_t)$ is known just for the selected $\alpha_t$.

Another alternative is to obtain theoretical regret bounds on the online learning problems over the original infinite set $\cala(\phi)$. The critical challenge in this setting for the regret analysis is to understand the properties of the per-round reward function. 
In particular, when the reward function is Lipschitz-continuous, sublinear regret algorithms can be shown in both the full information and bandit settings~\cite{maillard2010online}.
For the one-way trading problem (\otp), which could be interpreted as a simplified version of the knapsack problem, one can show the reward function under our proposed parametric algorithms is Lipschitz-continuous (See Appendix~\ref{app:reg-otp} for more detail). 

However, for \okp and \osc, the per-round reward functions are, in general, piecewise Lipschitz functions.
It is known that online learning algorithms for general piecewise Lipschitz functions suffer linear regret bounds~\cite{cohen2017online}, though recent work in~\cite{balcan2018dispersion} shows that sublinear regrets can be achieved if the piecewise Lipschitz reward functions satisfy some additional \textit{dispersion} conditions.
The key step of verifying the dispersion condition is to characterize the discontinuity locations of the reward function. 
For several classic offline problems, \cite{balcan2018dispersion} provided such characterizations for the reward functions corresponding to their parametric offline algorithms.
However, in the online learning problems of \okp and \osc, the reward functions are given by the optimal values of underlying online problems, and hence their discontinuities are challenging to characterize in general. However,
proving sublinear regret bounds by verifying the dispersion condition in specific application settings is promising, since in the OTP setting, the reward function can even be Lipschitz-continuous.

\section{Case Study}
\label{sec:exp} 

To illustrate the practical implications of being able to optimize within a policy class while still ensuring worst-case competitive bounds, we end the paper with a real-world case study  on the admission control of electric vehicles (EVs) in a charging station, which is an extended version of \okp. 


We consider a charging station with a charging demand more than its power capacity, which increasingly is the case. Upon the arrival of an EV, the station has to either admit or reject the request based on the  value and weight (amount of energy demand) of the charging  request as well as the current utilization of the station.  Clearly, this is similar to the \okp setting, with the difference being that the charging requests have flexibility within an available window and the stations may \textit{schedule} the charging requests with this flexibility. The station during the time can be seen as the multiple knapsacks.  Thus, the problem is a time-expanded version of single \okp. In \S\ref{app:ev} of the supplementary, we show the the relation between  EV-charging admission control problem and \okp more formally  and characterize its degradation factor rigorously. Also, we explain how to achieve the optimal competitive ratio of \okp for this extended problem. 



\textbf{Experimental setup.}  To explore the performance of our framework for EV admission control, we use  ACN-Data~\cite{LeeLiLow2019a}, an open EV dataset including over 50,000 fine-grained EV charging sessions. The data is collected from an EV parking lot at Caltech that includes more than 130 chargers. In this experiment, we use a two-month sequence of EV charging requests including arrival and departure times and charging demands. Since the data does not include the value of each request, we use a value estimation approach by modeling the distribution of historical arrivals and setting the values as a function of arrival, i.e., the higher the rate of arrival, the higher the value. Details are provided in \S\ref{app:exp} of supplementary. Moreover, we used a \textit{water-filling} scheduling policy to process the requests during the time. The water-filling policy splits the demand into the smaller parts then for each part, picks the slot with minimum utilization sequentially and updates the utilization by amount of smaller demand.  In the experiments, each instance considers one day, and we randomly generated 100 instances for each day, each with different values, and report the results for $60\times 100 = 6000$ instances.

We report the \textit{empirical profit ratio}, profit of optimal offline algorithm over the profit of online algorithm in experiment, of different algorithms, which is the counterpart of the theoretical competitive ratio in the empirical setting. We compare the empirical profit ratio of three different algorithms: (1) $\okpalg(1)$, the worst-case optimized online algorithm that does not take into account the power of learning from historical data, but, guarantees the optimal competitive ratio; (2) $\okpalgalphaoffline$, an algorithm that finds the best possible parameter in an offline manner. $\okpalgalphaoffline$ is not practical since it is fed with the optimal parameter; however, it illustrates the largest possible improvement from learning; and (3) $\okpalgalpha$, a simple, yet practical, algorithm that uses the optimal policy of the previous instance of the problem for the current instance. 


\begin{figure}[!t]
	\center
	\begin{minipage}[b]{.45\textwidth}
    	\includegraphics[width=0.98\textwidth]{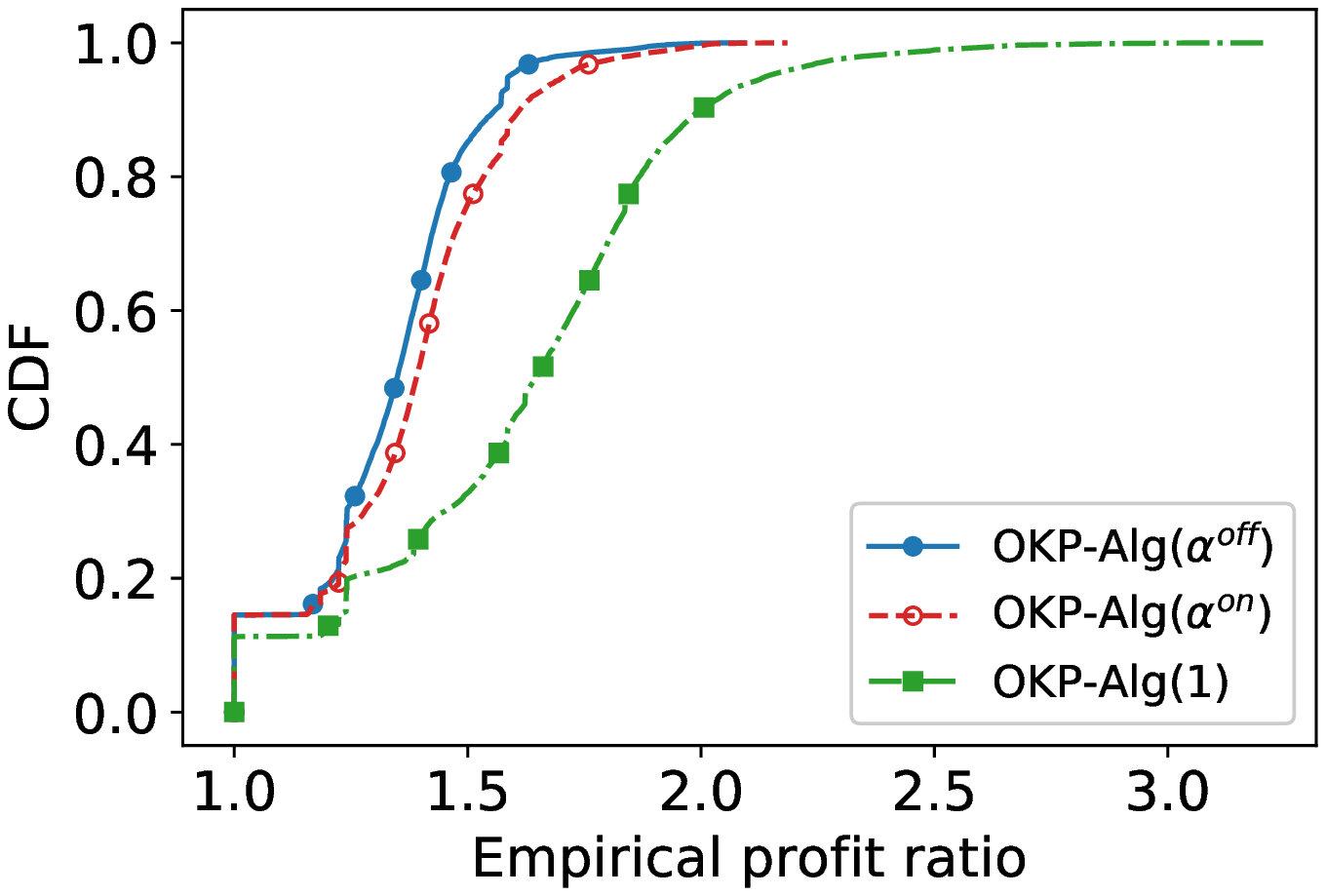}
        \vspace{-0.07in}
		\caption{CDF of empirical profit ratios of different algorithms solving \okp} 
		\label{fig:cdf}
	\end{minipage}\hspace{2mm}
	\begin{minipage}[b]{.45\textwidth}
	    \includegraphics[width=0.98\textwidth]{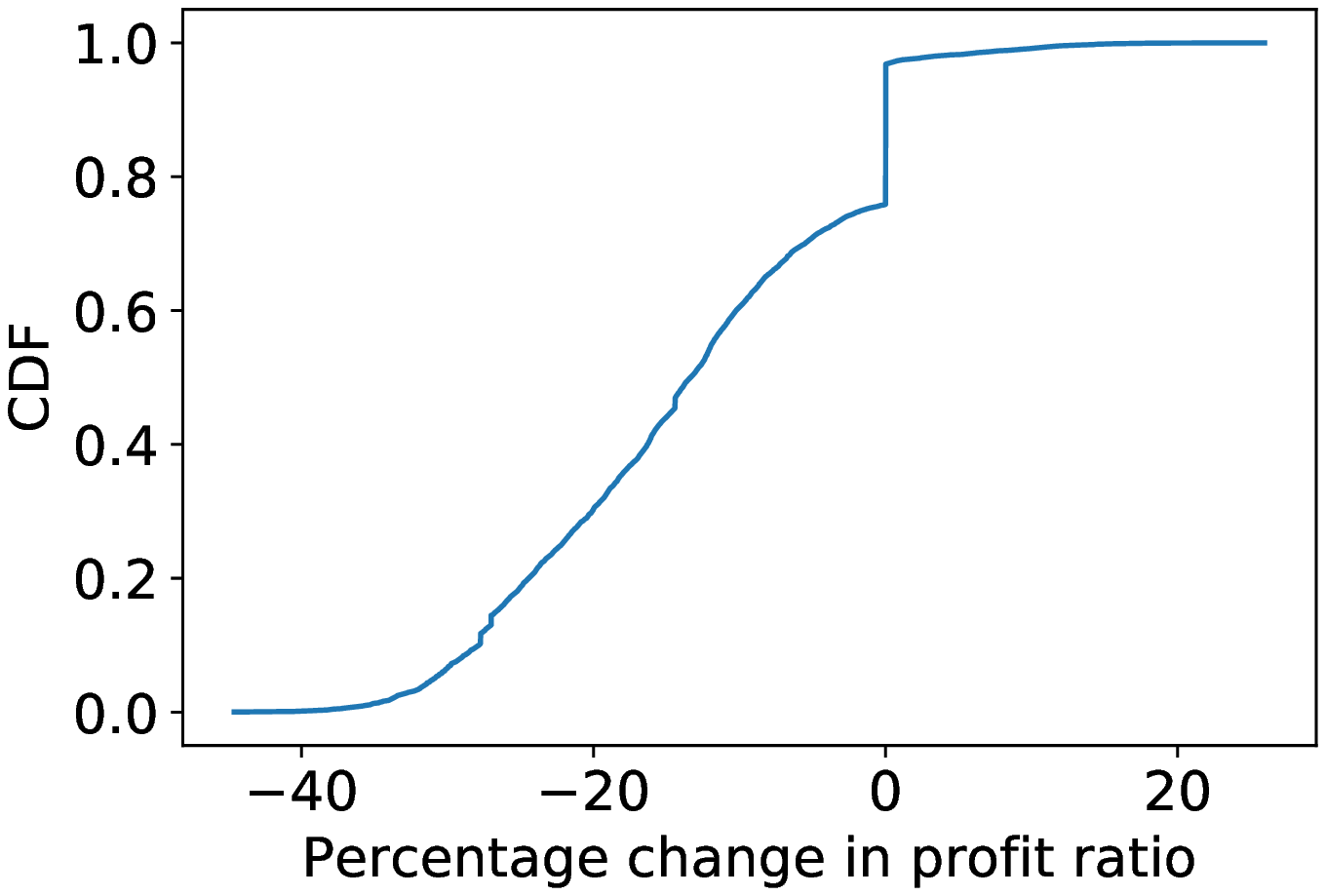}
	   \vspace{-0.07in}
		\caption{ Improvement  of $\okpalgalpha$ compared to $\okpalg(1)$} 
		\label{fig:cdf2}
	\end{minipage}
\end{figure}

\begin{figure}[!t]
	\center
	\begin{minipage}[b]{.45\textwidth}
    	\includegraphics[width=0.93\textwidth]{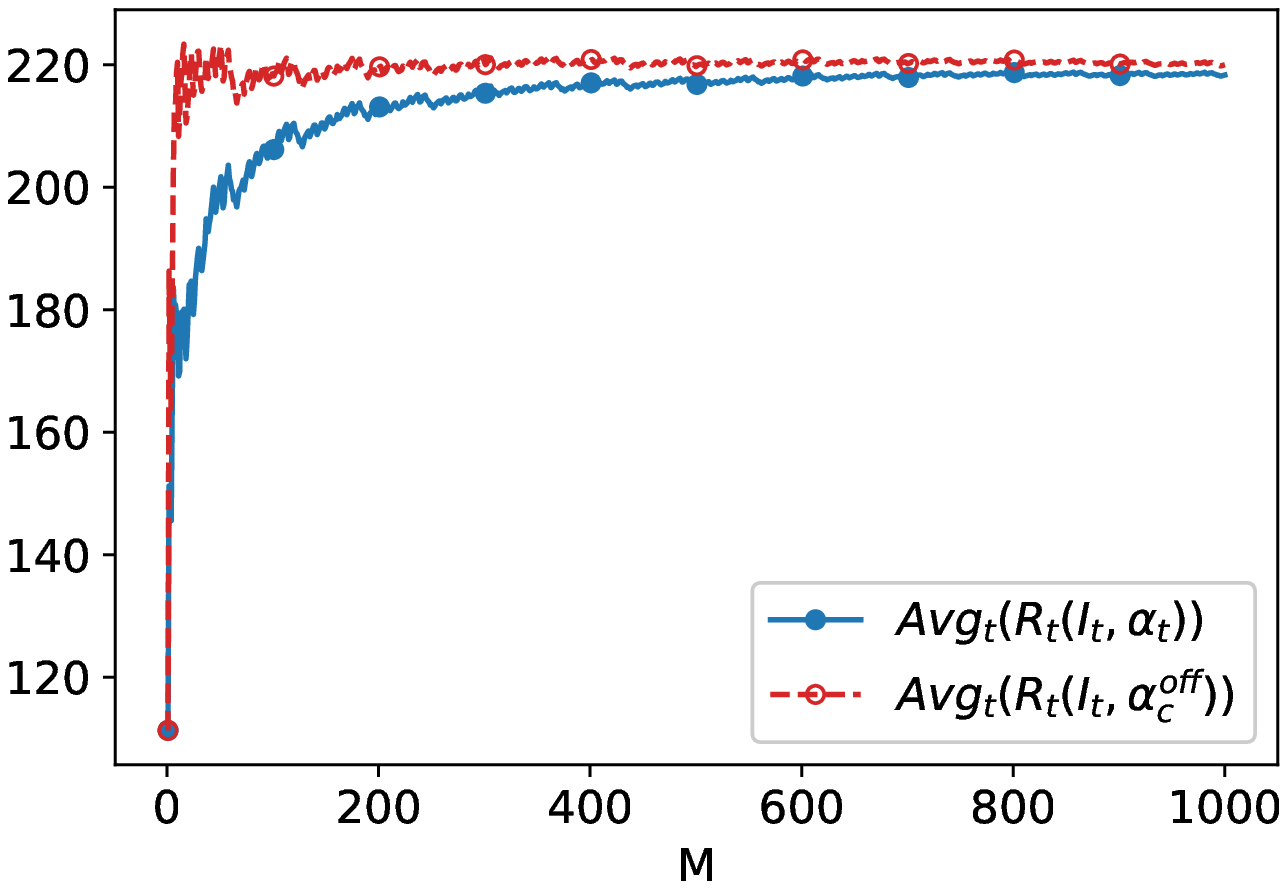}
		\caption{Average reward of the online learning algorithm and the static optimal} 
		\label{fig:regret_avg}
	\end{minipage}\hspace{2mm}
	\begin{minipage}[b]{.45\textwidth}
	    \includegraphics[width=0.999\textwidth]{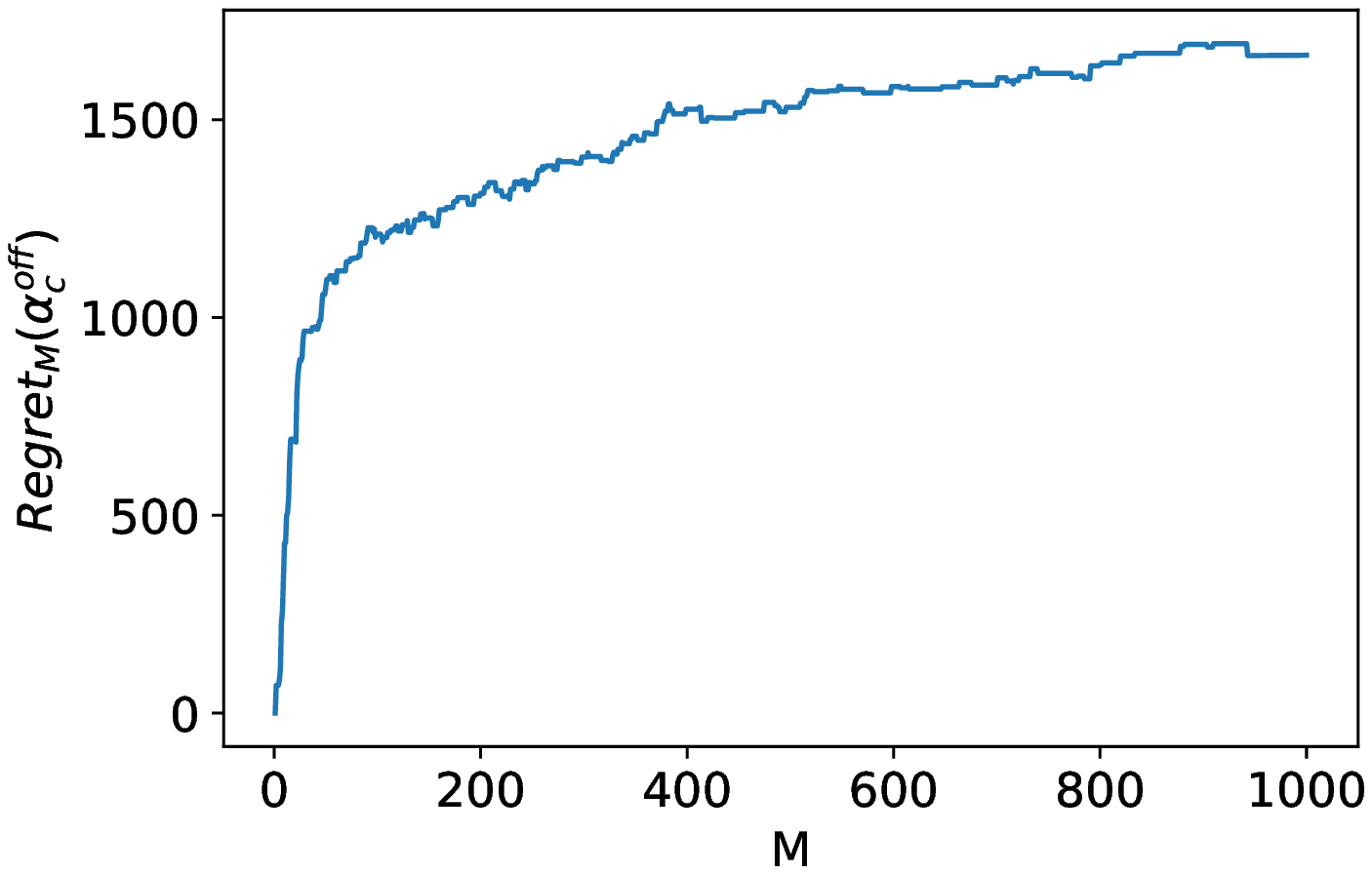}
		\caption{The regret of the online learning algorithm over different rounds} 
		\label{fig:regret_val}
	\end{minipage}
	\vspace{-0.4cm}
\end{figure}
 
\textbf{Experimental results.}
Figure~\ref{fig:cdf} plots the CDF of the empirical profit ratios of different algorithms. First, it shows that $\okpalgalphaoffline$, e.g., the offline optimal policy, substantially improves the performance, i.e., the 80th percentile of $\okpalg(1)$ has the profit ratio of 1.87, while with $\okpalgalphaoffline$, this is reduced to 1.46. Second, the performance of practical $\okpalgalpha$ is very close to that of $\okpalgalphaoffline$. To scrutinize the microscopic  behavior of $\okpalgalpha$, in Figure~\ref{fig:cdf2}, we plot the CDF of the improvement of the $\okpalgalpha$ as compared to $\okpalg(1)$. Notable observations are as follows: (i) approximately in 76\% of instances $\okpalgalpha$ outperforms $\okpalg(1)$, on average by 17.8\%; (ii) in 21\%, $\okpalgalpha$ has no benefits over $\okpalg(1)$; and finally, (iii) in  3\% of cases, $\okpalgalpha$ does worse than $\okpalg(1)$, on average by  6.4\%. A take-away from these experiments is that the average performance is substantially improved over the worst-case optimized algorithm, and that this improvement comes while only degrading the performance of a small fraction of instances by a small amount.

\textbf{Algorithm selection via online learning.}
In this experiment, we use an online learning approach for selecting the best algorithm (as described in Section \ref{sec:online_learning}). Specifically, we use the adversarial Lipschitz learning algorithm in a full-information setting to implement the parameter selection \cite{maillard2010online}. Figure~\ref{fig:regret_avg} reports the average reward collected by the online learning approach and the best offline static algorithm over different rounds of running the problem.  The average reward of an online algorithm converges to the optimal offline reward when as the learning process increases. The regret value's growth is presented in Figure~\ref{fig:regret_val}. When the average reward of the online algorithm merges to the optimal offline value, the marginal rate of regret value decreases although the regret value increases.  



\section{Concluding Remarks}
We developed an approach for characterizing policy classes of online algorithms with bounded competitive ratios, and introduce the \textit{degradation factor}, a new performance metric that determines the worst-case performance loss of learning the best policy as compared to a worst-case optimized algorithm. We apply our approach to the ski-rental, knapsack, and set cover problems. These applications serve as illustrations of an integrated approach for \textit{learning online algorithms}, while the majority of prior literature use ML for learning the uncertain input to online problems.

\section{Broader Impacts}
Our work fits within the broad direction of research on developing data-driven learning algorithms with guarantees. This is critically important for society, since data-driven algorithms are being increasingly adopted in safety-critical domains across sciences, businesses, and governments that impact people's daily lives. Our work proposes a framework that enables combining data-driven learning with the worst-case guarantees that traditional online algorithms provide, thus achieving the best of both worlds. That is, our approach leverages the power of ML-driven efficient decision making while adapting the safety guarantees from classic online algorithm design. We see no ethical concerns related to this paper. 

\section{Acknowledgment}
Ali Zeynali and Mohammad Hajiesmaili acknowledge the support from NSF grant CNS-1908298. Adam Wierman's research is supported by NSF AitF-1637598, and CNS-1518941. Also, Bo Sun received the support from Hong Kong General Research Fund, GRF 16211220.

 \bibliographystyle{unsrt}  
 \bibliography{references.bib}  
 \newpage
\appendix



\section{Related Work}
\label{app:rel_work}
In this section, we provide in-depth review of prior literature. We categorize the related literature into (i) prior literature on combining prediction (or advice) into the basic competitive framework (\S\ref{app:rel_1}); (ii) theory-driven approach for learning algorithms in a data-driven manner (\S\ref{app:rel_2}); and (iii) application-specific algorithm selection in practice (\S\ref{app:rel_3}). 

\subsection{Improving online algorithms with ML predictions}
\label{app:rel_1}
The prior literature on improving the practical performance of online algorithms can be divided into three categories: (i) traditional window-based predictions for online algorithms; (ii) online algorithms with advice; and more recently (iii) online algorithms with (untrusted) ML advice. In the following we briefly review this literature. 

The traditional window-based approach aims to feed the online algorithms with accurate or noisy inputs within a limited window of future time slots. This approach has been applied to a broad category of online problems such as ski-rental~\cite{lu2012simple}, energy scheduling~\cite{lu2013online}, and dynamic capacity provisioning of data centers~\cite{lin2012dynamic}. Following these initial papers, the approach has been  systematically extended to online convex optimization problems with switching cost in a sequence of results~\cite{chen2016using,chen2015online,comden2019online}. The underlying theoretical approach in this category provides elegant connections between model predictive control as well-established tool in the control theory and online algorithms. 


The second category looks at online algorithms with advice~\cite{boyar2017online}.
The idea behind this line of work is to measure how much information an online algorithms needs about the forthcoming online inputs to achieve a certain competitive ratio. This approach has been introduced in~\cite{bockenhauer2009advice,emek2011online} and extensively extended to a rich set of online problems including \okp~\cite{bockenhauer2014online} and \osc~\cite{komm2012advice}. For more details and problem that tackled in this domain, we refer to~\cite{boyar2017online}.

The last and most recent category aims to improve online algorithms with ML advice. This approach is introduced in~\cite{lykouris2018competitive}, with several followup works~\cite{angelopoulos2019online,purohit2018improving,rohatgi2020near,lattanzi2020online,gollapudi2019online}. The key motivation in this framework is two-fold: (1) keep the core competency of online algorithms, i.e., performance guarantee against worst-case; (2) achieve a provably improved performance if a good prediction is performed by the machine-learning tools. The first and second motivations are formally characterized by defining the notions of robustness and consistency. Robustness measures how well the algorithm does in the worst-case of bad prediction, and consistency measures how well the algorithm does under perfect prediction. This method can also be expanded to the setting of ML advice from multiple experts, e.g., ~\cite{gollapudi2019online}, and provides some elegant tradeoffs in the case when the expert advice is conflicting.

However, in above literature, the power of ML models has been leveraged to first, predict the future input, and then modify the algorithms to use this additional input to further improve the performance. In this way, the learning and algorithmic parts are decoupled, and thus worst-case performance guarantees only hold when the predictions are (nearly) perfect. In practice, however, it is not always possible to predict the fine-grained future input. Rather, in most application domains, it is possible to learn some useful coarse-grained information (see for example our intuition for \osc in Figure~\ref{fig:osc_intuition}) that could be leveraged in algorithm design. In other words, engineers typically prefer to optimize over a policy class of parametrized algorithms and then tune the parameters using a historical set of problem instances from their domain to find a policy with high expected performance over future problem instances. The current literature cannot provide a design space for learning online algorithms from a policy class. To fill this gap, this paper proposes an integrated approach for developing adaptive (data-driven) online algorithms with worst-case guarantees. This approach has recently been advocated for in theoretical computer science community, as explained in the following. 




\subsection{Data-driven Algorithm Design}
\label{app:rel_2}
The high-level idea of our work on defining a policy class of online algorithms and learning the best policy within the class is in the same spirit as the emerging direction of data-driven algorithms~\cite{gupta2020data}. This direction was introduced in a seminal paper of Gupta and Roughgarden~\cite{gupta2017pac}, and has received considerable attention since, e.g.,~\cite{kleinberg2019procrastinating,balcan2018dispersion,alabi2019learning,cohen2017online}. See~\cite{gupta2020data} for more references. While the motivation of our work is the same as these works, our focus is on data-driven \emph{competitive} algorithms. Specifically, the framework introduced in~\cite{gupta2017pac} is a general one for both offline (batch) and online settings. For the online setting, the focus is on regret minimization strategies for selecting the best algorithm. Our work is the first to bring the idea of learning the best online algorithm among a policy class of algorithms to the setting of competitive analysis. In addition, we introduce degradation factor, a new performance metric that can explicitly characterize the worst-case performance loss due to the data-driven learning. 


\subsection{Empirical Parameter Tuning of Online Algorithms}
\label{app:rel_3}
The is a large literature on empirical approaches for parameter tuning and more broadly data-driven algorithm design, see~\cite{gupta2020data} and references there in for examples. For online algorithms, in particular, a recent work~\cite{akhtar2018oboe} proposes an empirical parameter tuning approach for adaptive bitrate streaming algorithms, a celebrated video streaming problem. 
The substantial practical improvement in these empirical studies demonstrate the benefits of our approach, as we also show in our case study. In contrast to our work, the empirical approach does not provide theoretical understanding of the data-driven algorithm selection. 
Last it is worth noting that to perform the optimization in practice, several tools like \textit{Google Vizier} \cite{golovin2017google} could be used to provide a service for black-box optimization and parameter tuning.


\section{Proof of Theorem \ref{thm:cr_ski}}
\label{app:ski}
We consider two cases: (i) $b < p$, in this case, if the actual number of skiing days is no larger than $b$, the competitive ratio of $A(b)$ is 1. If the number of skiing days is $x > b$, the competitive ratio of algorithm is
$$\max_x \{ \frac{b+p}{\min(x, p)} \} = \frac{b + p}{b + 1} < \frac{b + p}{b} = 1 + \frac{p}{b}.$$

(ii) When $b \ge p$, if actual number of skiing days is less than $p$, algorithm is optimal. Otherwise, the competitive ratio of $A(b)$ is
$(b + p)/{p} = 1 + b/p$. Putting together (i) and (ii), we have
$$    \texttt{CR}(A(b)) = 1 + \max \{p/b,b/p\}.$$
Finally, the degradation factor of $A(b)$ with respect to base line algorithm $A(p)$ would be:
\begin{eqnarray*}
\texttt{DF}(A(b)) &=& \frac{\texttt{CR}(A(b))}{\texttt{CR}(A(p))} = \frac{\texttt{CR}(A(b))}{2} \\
&=& 1/2 + \max \{p/2b,b/2p\}.
\end{eqnarray*}

\section{The Online Knapsack Problem and One-Way Trading Problem}
\label{app:okp_otp}
The online knapsack problem (\okp) is closely related to the one-way trading problem (\otp)~\cite{el2001optimal,Lin2019Sigmetrics} in the sense that the optimal online algorithms for \okp and \otp are both threshold-based algorithms parameterized by the same threshold function, and achieve the same optimal competitive ratio. In \otp, a seller aims to sell one unit of an infinitely-divisible item to a set $[n]$ of buyers that arrive sequentially. Upon the arrival of buyer $i$, the seller must decide the fraction of the item, $x_i \geq 0$, to sell, and correspondingly can get a revenue $g_i(x_i)$, which is a concave non-decreasing function.
The seller has no ideas on the revenue functions and the total number of buyers.
The goal of \otp is to maximize the total revenue from all buyers. 
The offline \otp can be formulated as below:
\begin{subequations}
\label{eq:OTP_form}
    \begin{eqnarray*}
      [\textrm{Offline}\ \otp] \quad \max & \sum_{i\in[n]}g_i(x_i), \\ 
      \textrm{s.t.,}&  \quad \sum_{i\in[n]} x_i\leq 1, \\ 
      \textrm{vars.,}& \quad  x_i \ge 0, \quad i \in [n],
    \end{eqnarray*}
\end{subequations}
where the revenue function is assumed to have bounded marginal revenues, i.e., $g_i'(x_i) \in [L,U],\forall i\in[n]$.
From the offline formulations, it can be noted that \otp is a continuous version of \okp with a more general revenue function.
When $g_i(x_i) = v_i x_i$ takes a linear form, \otp reduces to the classic one-way trading problem that was introduced in~\cite{el2001optimal}. 

Similar to the online algorithm for \okp, the algorithm for \otp also uses the idea of threshold-based decision making. In contrast to the binary decision in \okp, the \otp determines a continuous fraction of an item for selling to each buyer. 
Given a threshold function $\Psi_\alpha(z)$, the online algorithm for \otp, $\otpalg(\alpha)$, is a modified version of Algorithm~\ref{alg:ota} with continuous decisions. In particular, upon the arrival of buyer $i$ with a revenue function $g_i(\cdot)$, the online algorithm determines the amount of the selling item by solving another pseudo-utility maximization problem
\begin{align}\label{p:otp}
x_i^* = \argmax_{x_i\ge0} g_i(x_i) - \int_{z_{i-1}}^{z_{i-1}+x_i}\Psi_\alpha(u)du,
\end{align}
where $\int_{z_{i-1}}^{z_{i-1}+x_i}\Psi_\alpha(u)du$
estimates the cost of $x_i$ fraction of the item when current utilization is $z_{i-1}$. The utilization is then updated by
\begin{align}\label{eq:otp-utilization}
z_{i} = z_{i-1} + x_i^*.
\end{align}
$\otpalg(\alpha)$ can achieve the optimal competitive ratio for \otp, $\ln(\gamma) + 1$, when the threshold function is the same as that of \okp, i.e., $\Psi_{\alpha}$ in Equation~\eqref{eq:thresholdf_prime} with $\alpha = 1$. 
To sum up, we can construct the policy class of \otp as $\otpalg(\alpha)$ by replacing the pseudo-utility maximization problem and utilization update in Lines~3~and~4 of Algorithm~\ref{alg:ota} with Equations~\eqref{p:otp}~and~\eqref{eq:otp-utilization} and keeping the same parametric threshold function $\Psi_\alpha$.   
Since the online algorithms $\okpalg(\alpha)$ and $\otpalg(\alpha)$ are parametrized by the same threshold function $\Psi_\alpha$ and their competitive ratios can be expressed as the same function of those parameters
, our results for \okp naturally hold for \otp.

\section{Proof and Remarks on Theorem~\ref{thm:DF_OKP}}
\label{app:okp}

\subsection{Proof of Theorem~\ref{thm:DF_OKP}}

We start with the competitive analysis of the optimal online algorithm $\okpalg(1)$ for \okp.
Let $z^* := z_n$ denote the final utilization of the knapsack after processing all $n$ items. Also let $\mathcal{I}(z^*)$ denote the set of instances, which result in utilization $z^*$ after running $\okpalg(1)$.
The values of the online algorithm under the worst-case instance in $\mathcal{I}(z^*)$ and its corresponding offline optimum are denoted by ${\sf ALG}(z^*)$ and ${\sf OPT}(z^*)$, respectively.
For each $z^*\in[0,1]$, the worst-case instance in $\mathcal{I}(z^*)$ can be constructed as follows. 

When $z^*\in[0,T]$, i.e., the final utilization falls in the flat segment of the threshold function, the worst case occurs when all items have the same value density $L$ and their total weights are $z^*$. In this case both online algorithm and the offline optimum will admit all items and achieve the same total value. Thus, when $z^*\in[0,T]$, the competitive ratio is ${\sf OPT}(z^*)/{\sf ALG}(z^*) = 1$. 

When $z^*\in(T,1]$, all $n$ items in the worst-case instance are divided into two batches. The items in the first batch are indexed by $i\in[n_1]$. The weights of the items in this batch satisfy $\sum_{i\in[n_1]}w_i = z^*$ and the value densities of the items are given by $v_i/w_i = \Psi_1(\sum_{j\in[i-1]}w_{j}), i\in[n_1]$. The items in the second batch are indexed by $i\in[n_2]$. All items in this group have the same value density $\Psi_1(z^*) - \epsilon$, where $\epsilon$ is an arbitrarily small positive number, and their total weights can occupy the whole capacity, i.e., $\sum_{i\in[n_2]}w_i = 1$. Under this instance, the online algorithm $\okpalg(1)$ only takes the first batch of the items and reject all those in the second batch. The total value of the online algorithm is then $\alg(z^*) = \sum_{i\in[n_1]}w_i \Psi_1(\sum_{j\in[i-1]}w_{j}) \approx \int_{0}^{z^*} \Psi_1(u)du$ since $w_i \ll 1$. While the offline optimum will only admit the items in the second batch and achieves the total value $\opt(z^*) = \Psi_1(z^*) - \epsilon$.  
Then the competitive ratio of $\okpalg(1)$ is \begin{align}\label{eq:okp-cr}
\max_{z^*\in[0,1]}\max_{\textrm{instances\ in}\ \mathcal{I}(z^*)} \frac{\opt(z^*)}{\alg(z^*)}  \\ = \max_{z^*\in(T,1]} \frac{\Psi_1(z^*)}{\int_{0}^{z^*} \Psi_1(u) du} = \ln(\gamma) + 1.
\end{align}

We next show the degradation factor of $\okpalg(\alpha), \alpha>0,$ with respect to $\okpalg(1)$. 
For notation convenience, let $\texttt{DF}_\alpha := \texttt{DF} \left(\okpalg(\alpha)\right)$. When $z^* \in [0,T]$, the worst-case instance of $\okpalg(\alpha)$ is the same as that of $\okpalg(1)$, and this case will not affect the derivation of degradation factor. Therefore, we focus on the instances that result in $z^* \in (T,1]$. 

When $\alpha \ge 1$, the worst-case instance of $\okpalg(\alpha)$ is also the same as that of $\okpalg(1)$ after replacing $\Psi_1$ with $\Psi_\alpha$ when setting the value densities of the items.
Under this worst-case instance, based on the definition of the degradation factor and substitution of $\opt(z^*)$ and $\alg(z^*)$, we have
\begin{align}\label{eq:DF-proof1}
\Psi_{\alpha}(z^*) - \epsilon \le \frac{\texttt{DF}_\alpha}{T} \int_0^{z^*} \Psi_{\alpha}(z) dz, \quad z^*\in(T,1],
\end{align}
where $1/T = \ln(\gamma)+1$ is the competitive ratio of the baseline algorithm $\okpalg(1)$.

Let $z^{\texttt{u}}$ denote the utilization level at which the threshold function $\Psi_\alpha$ reaches the maximum value density, i.e., $\Psi_\alpha(z^{\texttt{u}}) = U$. 
We substitute $\Psi_\alpha$ into Equation \eqref{eq:DF-proof1} and show the degradation factor $\texttt{DF}_\alpha$ in the following two sub-cases: (i-a) $\alpha \geq 1$ and $T< z^* \leq z^{\texttt{u}}$, (i-b) $\alpha \geq 1$ and $z^{\texttt{u}} < z^*\le 1$. 

\textbf{Case (i-a): $\alpha \ge 1$ and $T< z^* \leq z^{\texttt{u}}$:}
\begin{eqnarray*}
L e^{\alpha z^*/T -\alpha} - \epsilon &=& \frac{\texttt{DF}_\alpha}{T} \int_0^{z^*} \Psi_{\alpha}(z) dz,\\ 
&=& \frac{\texttt{DF}_\alpha}{T} \bigg[LT + \int_T^{z^*}Le^{ \alpha z/T - \alpha} dz \bigg],\\
&=& \frac{\texttt{DF}_\alpha}{T} \bigg[LT + \frac{LT e^{-\alpha}}{\alpha} [e^{ \alpha z^*/T} - e^{\alpha}] \bigg],\\
&=& \texttt{DF}_\alpha \bigg[L + \frac{L}{\alpha} [e^{-\alpha} e^{\alpha z^*/T} - 1]\bigg].
\end{eqnarray*}
Therefore, in Case (i-a), $\texttt{DF}_\alpha$ is upper bounded by
\begin{equation}
    \texttt{DF}_\alpha = \frac{\alpha e^{-\alpha} e^{\alpha z^*/T}}{\alpha + e^{-\alpha} e^{\alpha z^*/T} - 1} \leq \frac{\alpha \gamma }{\alpha +  \gamma - 1}.
\end{equation}
\textbf{Case (i-b): $\alpha\ge 1$ and $z^{\texttt{u}} < z^* \le 1$:}
\begin{eqnarray*}
U - \epsilon &=& \frac{\texttt{DF}_\alpha }{T} \int_0^{z^*} \Psi_{\alpha}(z) dz,\\
&=& \frac{\texttt{DF}_\alpha}{T} \bigg[LT + \int_T^{z^{\texttt{u}}}Le^{\alpha z/T - \alpha} dz + (z^* - z^{\texttt{u}})U\bigg],\\
&=& \frac{\texttt{DF}_\alpha }{T} \bigg[LT + \frac{L Te^{-\alpha}}{\alpha} [e^{\alpha z^{\texttt{u}}/T} - e^{\alpha}] + (z^*- z^{\texttt{u}})U \bigg],\\ &=& \texttt{DF}_\alpha  \bigg[L + \frac{L}{\alpha} [e^{-\alpha} e^{\alpha z^{\texttt{u}}/T} - 1] + (z^*- z^{\texttt{u}}) U/T \bigg].
\end{eqnarray*}
We then have
\begin{equation}
     \texttt{DF}_\alpha  = \frac{U \alpha}{L \alpha + U - L + (z^* - z^{\texttt{u}}) \alpha U/T} \leq \frac{\alpha \gamma }{\alpha +  \gamma - 1}.
\end{equation}
Combining Cases (i-a) and (i-b) gives the degradation factor $\texttt{DF}_\alpha = \frac{\alpha \gamma }{\alpha +  \gamma - 1}$, when $\alpha \ge 1$. 

When $0<\alpha < 1$, the worse-case instance is changed. First, the total weights of the items in the first batch are $1$ and their values are still $v_i = \Psi_\alpha(\sum_{j\in[i-1]}w_{j}), i\in[n_1]$.  
Second, the values of the items in the second batch are all equal to $U$ and their total weights are still $1$. Under this worst case, the degradation factor can be characterized as follows.
\begin{align}\label{eq:DF-proof}
U \le \frac{\texttt{DF}_\alpha}{T} \int_0^{1} \Psi_{\alpha}(z) dz, z^*\in(T,1].
\end{align}
Thus, we can derive the degradation factor from the following case.

\textbf{Case (ii): $0< \alpha < 1$:}
\begin{eqnarray*}
U &=& \frac{\texttt{DF}_\alpha}{T} \bigg[ LT + \frac{LT e^{-\alpha}}{\alpha } [ e^{\alpha/T} - e^{\alpha}] \bigg],\\
&=& \texttt{DF}_\alpha \bigg[ L + \frac{L e^{-\alpha}}{\alpha}[e^{\alpha /T} - e^{\alpha}] \bigg].
\end{eqnarray*}
Thus, we have
\begin{equation}
    \texttt{DF}_\alpha = \frac{\alpha \gamma}{\alpha + e^{-\alpha} e^{\alpha/T} - 1} = \frac{\alpha \gamma}{\alpha + \gamma ^{\alpha} - 1},
\end{equation}
where we use $e^{1/T-1} = \gamma$. Thus, the degradation factor $\texttt{DF}_\alpha = \frac{\alpha \gamma}{\alpha + \gamma ^{\alpha} - 1}$, when $0 < \alpha < 1$.


\subsection{Remarks on Parameter Sensitivity of Degradation Factors}
\begin{figure}
\begin{minipage}[b]{0.45\textwidth}
    \centering\includegraphics[width=0.95\linewidth]{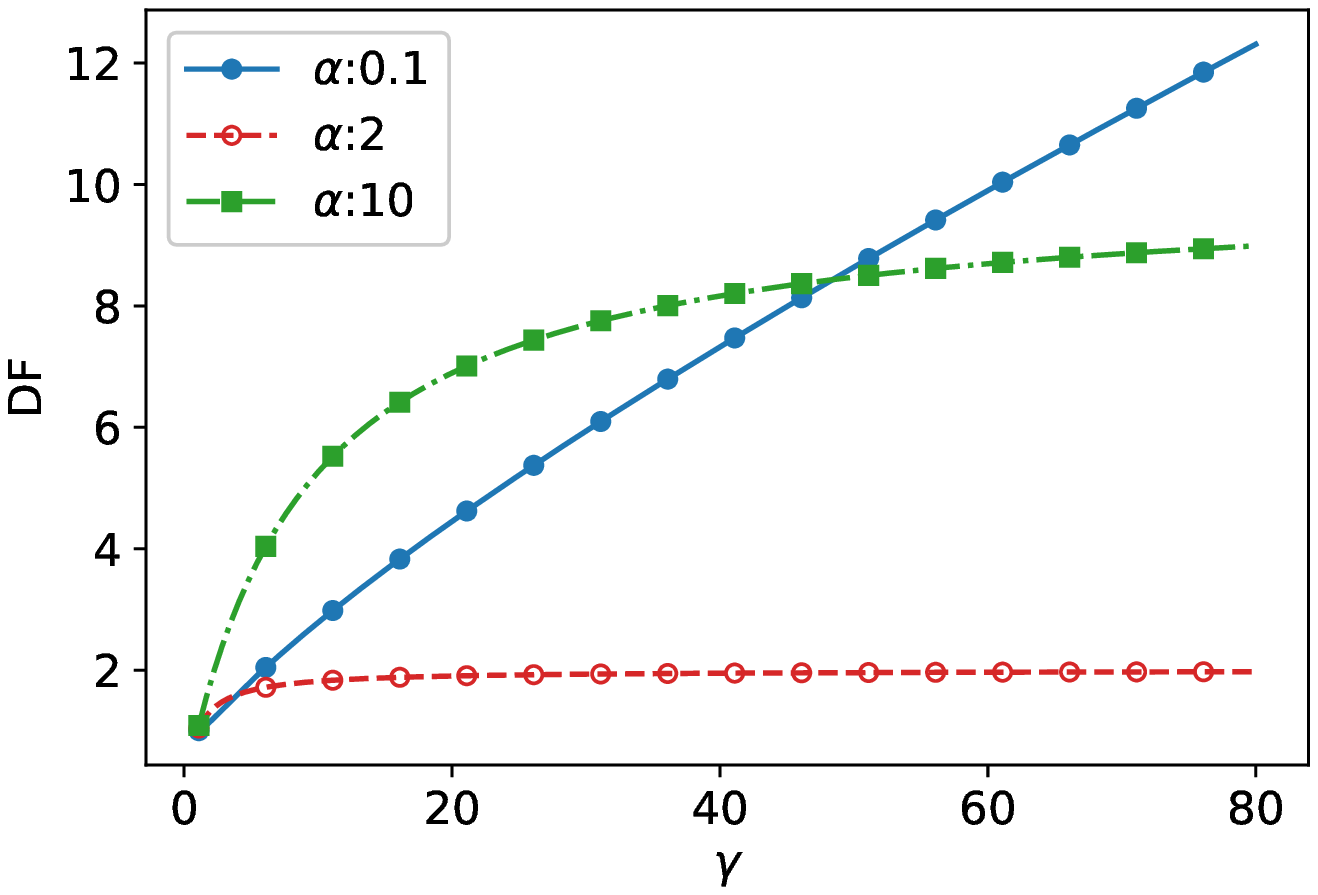}
    \caption{\texttt{DF} vs. $\gamma$ for different values of  $\alpha$}
    \label{fig:okp_beta_gama_0.1}
\end{minipage}\hspace{0mm}
\begin{minipage}[b]{0.45\textwidth}
    \centering\includegraphics[width=0.95\linewidth]{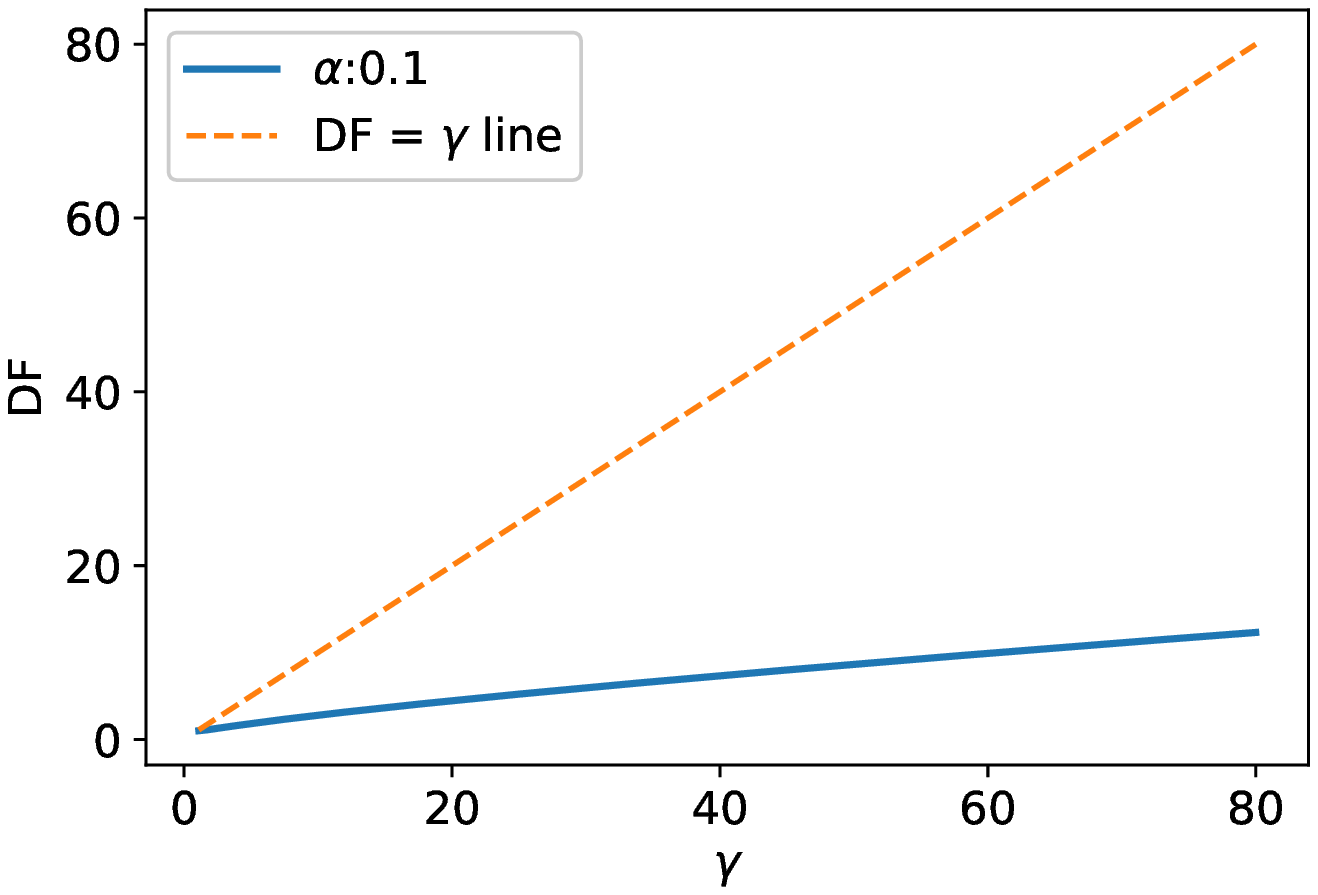}
    \caption{\texttt{DF} vs. $\gamma$ for $\alpha = 0.1$}
    \label{fig:okp_beta_gama_10}
\end{minipage}\hspace{2mm}

\end{figure}

By characterizing the degradation factor of \okp with respect to \okpalg(1), \texttt{DF}$\left(\okpalg(\alpha)\right)$,  
we are interested in how it changes as a function of $\alpha$ as well as additional parameters of the problem, e.g., the fluctuation ratio $\gamma$.
As shown in Figure~\ref{fig:price-of-learning}, \texttt{DF}$\left(\okpalg(\alpha)\right)$ is minimized at $\alpha = 1$, and grows when $\alpha$ deviates from $1$. Note that the degradation factor is very sensitive to the changes of $\alpha$ when $\alpha \in (0,1)$ while it changes at a sub-linear rate when $\alpha \in (1,+\infty)$. This phase change originates from the changes of the worst-case scenarios when $\alpha$ is tuned to $(0,1)$ and $(1,+\infty)$, respectively.
In Figures \ref{fig:okp_beta_gama_0.1} we further depict the growth of the degradation factor of $\okpalg(\alpha)$ as a function of  $\gamma$  for different values of $\alpha$. The result shows that when $\alpha < 1$ the degradation factor grows faster than $\alpha >1$. However, the growth of degradation factor for $\alpha<1$ is still sublinear.    To see this, in Figure \ref{fig:okp_beta_gama_10} the value of degradation factor as a function of $\gamma$ for the especial case $\alpha = 0.1$ is compared to the linear growth. The result shows that the degradation factor also degrades gracefully as $\gamma$ increases at a sublinear rate. 

\section{Proof of Theorem \ref{thm:df_osc}}
\label{app:osc}
In the following, we prove the result in Theorem~\ref{thm:df_osc}, which characterizes the degradation factor \texttt{DF}$(\oscalg(\theta))$ for the \osc problem. The proof is inspired by techniques used in~\cite{alon2003online}, however, it additionally accounts for the feasibility of the steps once the parameter $\theta$ is adaptive. The proof consists of three steps. First, we show that the total number of weight-augmentation steps is bounded. Second, we prove that it is feasible to select some subsets such that the potential function does not increase, i.e., the procedure in Line~\ref{algline:update} of \oscalg is feasible. Finally, by combining these results, we characterize the degradation factor of $\oscalg(\theta)$ with respect to the baseline algorithm $\oscalg(2)$.

In the first step, we prove that the total number of weight-augmentation steps is at most ($2+ \frac{\log(m)}{\log(\theta)}$). During the execution of \oscalg, for any subset $s\in\mathcal{S}$, $w_s$ is no larger than $\theta$. During each weight-augmentation step, for each $s\in\mathcal{S}_i$, $w_s$ is multiplied by $(\theta)^k$. So, to ensure that $w_s \leq2$, the number of weight-augmentation steps is at most $\log_{\theta}{\big( \theta^2 m \big)} = 2 + \frac{\log m}{
\log (\theta)}$.

In the second step, we prove the feasibility of selecting some subsets while ensuring the potential function does not increase. More formally, following the statement in Line~\ref{algline:update} of \oscalg, we prove that when element $i$ arrives, and $i$ is not covered yet, there is a choice of at most $2 \theta \log n$ among sets in $\mathcal{S}_i$ such that after adding these sets to  $\mathcal{S}^{\texttt{sel}}$, the value of potential function does not increase.

To prove this, let $\Phi$ and $\Phi'$ be the values of the potential function before and after executing Line~\ref{algline:update} for covering element $i$. Also, for each $s\in\mathcal{S}_i$, let $w_s' = w_s + \delta_s$ be  the updated weight of set $s$ after weight-augmentation step. Similarly, let $w_i'$ be the updated weight of element $i$ while the difference with previous weight is $\delta_i$, i.e., $w_i' = w_i + \delta_i$. 
Now, we proceed with the selection procedure as follows. Consider an iterative process for selecting $2\theta \log n$ subsets in $\mathcal{S}_i$. In each iteration, we pick at most one subset, $s$ with probability $\delta_s/\theta \leq 1$ for picking $s$.

Consider element $j\notin \mathcal{I}^{\texttt{sel}}$, let $n_{i,j}$ be number of selected subsets upon arrival of $i$ containing $j$. So, the contribution of $j$ in potential function, denoted by $\Phi_j, j\notin \mathcal{I}^{\texttt{sel}}$, is
\begin{align}
\Phi_j = 
    \begin{cases}
    n^{2w_j'}, & \text{if }  n_{i,j} = 0,\\
    0, & \text{otherwise}.
    \end{cases}
\end{align}
Given the number of iterations be $2\theta \log n$ and $1-\delta_j/\theta$ be the probability of not covering $j$ in each iteration, the expected value of $\Phi_j$ is $(1 - \frac{\delta_j}{\theta})^{2\theta \log n}( n^{2w_j'})$. We can evaluate the upper-bound for this value:
\begin{equation*}
    (1 - \frac{\delta_j}{\theta})^{2\theta \log n}( n^{2w_j'}) \leq (n^{-\frac{2\theta \delta_j }{\theta} })(n^{2w_j + 2\delta_j}) = (n^{2w_j}).
\end{equation*}
Therefore, $\mathbb{E}[\Phi'] \leq \Phi$. This means that, there exist a choice of at most $2 \theta \log n$ such that $\Phi' \leq \Phi$.



The offline optimal solution to \osc problem includes at least one set. Hence, combining the results in the first and second steps, the competitive ratio of $\oscalg(\theta)$ is
\begin{equation}
    \label{eq:CR_OSC}
    \texttt{CR}(\oscalg(\theta)) = (2 \theta \log n)(2 + \frac{\log m}{\log \theta}).
\end{equation}
Last, the degradation factor of \texttt{$\oscalg(\theta)$} with respect to \texttt{$\oscalg(2)$} would be:
\begin{eqnarray*}
    \texttt{DF}(\oscalg(\theta)) &=& \frac{\texttt{CR}(\oscalg(\theta))}{\texttt{CR}(\oscalg(2))}\\& =& \frac{(2 \theta \log n)(2 + \frac{\log m}{\log \theta})}{(4 \log n)(\log m + 2)}
     \\ &=& \theta \bigg[ \frac{2\log \theta + \log m}{2\log \theta (2+\log m)} \bigg].
\end{eqnarray*}
This completes the proof. It is worth noting that with the closed-form competitive ratio in Eq.~\eqref{eq:CR_OSC}, one may find the minimum value of competitive ratio as the worst-case optimized one. And interestingly, this value is a function of $m$, as the number of subsets, and $\theta=2$, as the algorithm proposed in~\cite{alon2003online} is a suboptimal algorithm. This is indeed our observation in Figure~\ref{fig:osc_beta_theta}, where the degradation factor for $m=10^5$ with $\theta=[2,3]$ is less than 1. This shows that the original algorithm proposed in~\cite{alon2003online} is not using the worst-case optimized parameter and one can find the worst-case optimized parameter $\theta$ by minimizing Eq.~\eqref{eq:CR_OSC}.

\section{Regret Analysis of Data-driven Algorithm Design for One-Way Trading Problem}
\label{app:reg-otp}

We show that the reward function of \otp (see Appendix~\ref{app:okp_otp} for more detail) can be derived as closed-form Lipschitz continuous functions and its corresponding online learning problem can be solved with a sublinear regret.

We consider the classic \otp setting, in which the revenue function is linear and an instance of \otp is a sequence of exchange rates (or prices). 
Let $v_i^{(t)}$ denote the $i$-th exchange rate in round $t$.
Given an instance $\cali_t=\{v_1^{(t)},\dots,v_n^{(t)}\}$ of \otp
in round $t$, we characterize the reward function $R_t(\cali_t,\alpha)$ as a function of the tuning parameter $\alpha$.
Since the non-increasing exchange rates in $\cali_t$ can be eliminated without affecting the reward of round $t$, it is sufficient to focus on the instances with strictly-increasing exchange rates, i.e., $L\le v_1^{(t)} <... < v_n^{(t)}\le U$.
The reward function $R_t(\cali_t,\alpha)$ can be derived as follows.

Based on the online algorithm $\otpalg(\alpha)$, which is described in Appendix~\ref{app:okp_otp}, with the threshold function $\Psi_\alpha(z)$, the online traded resource of \otp can be determined as
\begin{align*}
    x_i^{(t)}=
    \begin{cases}
    \Phi_\alpha(v_1^{(t)}) & i = 1\\
    \Phi_\alpha(v_{i}^{(t)}) - \Phi_\alpha(v_{i-1}^{(t)}) & i=2,\dots,n
    \end{cases},
\end{align*}
where $\Phi_\alpha$ is the inverse function of the threshold $\Psi_\alpha$. 



The reward function of \otp is divided into multiple segments over $\alpha\in(0,+\infty)$ and the closed-form expressions of each segment can be derived as follows.
Depending on the maximum exchange rate $v_n^{(t)}$ in an instance, the reward function can be in two different forms.

\noindent\textbf{Case (i):} the maximum rate is lower than $U$, i.e., $v_n^{(t)} < U$.

When $\alpha \in(0, \frac{\ln(v_1^{(t)}/L)}{\ln(U/L)}]$, we have $\Phi_\alpha(v_i^{(t)}) = 1, \forall i\in[n]$ and hence $x_1^{(t)} = 1$ and $x_i^{(t)} = 0, i=2,\dots,n$.
Thus, the reward function is a constant
\begin{align*}
    R_t(\cali_t,\alpha) = v_1^{(t)}. 
\end{align*}

When $\alpha \in (\frac{\ln(v_{i-1}^{(t)}/L)}{\ln(U/L)},\frac{\ln(v_i^{(t)}/L)}{\ln(U/L)}], i=2,\dots,n$, we have
\begin{align*}
    &R_t(\cali_t,\alpha) = \sum_{i\in[n]} v_i^{(t)} x_i^{(t)}\\
    &= v_1^{(t)} \Phi_{\alpha}(v_1^{(t)}) + \sum_{j=2}^{i-1}  v_j^{(t)}[\Phi_{\alpha}(v_j^{(t)}) - \Phi_{\alpha}(v_{j-1}^{(t)})] + v_i^{(t)}[1 - \Phi_{\alpha}(v_{i-1}^{(t)})]\\
    &:= a_i^{(t)} - \frac{b_i^{(t)}}{\alpha},
\end{align*}
where $a_i^{(t)} = v_i^{(t)} + ( v_1^{(t)} -  v_i^{(t)})T$, $b_i^{(t)}=\sum_{j=1}^{i-1} T(v_{j+1}^{(t)} - v_j^{(t)})\ln(v_j^{(t)}/L)$. 

When $\alpha \in (\frac{\ln(v_n^{(t)}/L)}{\ln(U/L)},+\infty)$, we have 
\begin{align*}
    &R_t(\cali_t,\alpha) = \sum_{i\in[n]} v_i^{(t)} x_i^{(t)}\\
    &= v_1^{(t)} \Phi_{\alpha}(v_1^{(t)}) + \sum_{i=2}^{n}  v_i^{(t)}[\Phi_{\alpha}(v_i^{(t)}) - \Phi_{\alpha}(v_{i-1}^{(t)})]\\
    &= v_1^{(t)} T + \frac{1}{\alpha}\sum_{i\in[n]}T\ln(v_i^{(t)}/v_{i-1}^{(t)})\quad (v_0^{(t)} := L)\\
    &:= c^{(t)} + \frac{d_i^{(t)}}{\alpha},
\end{align*}
where $c^{(t)} = v_1^{(t)} T$, and $d_i^{(t)} = \sum_{i\in[n]}T\ln(v_i^{(t)}/v_{i-1}^{(t)})$.


\noindent\textbf{Case (ii):} the maximum rate is $U$, i.e., $v_n^{(t)} = U$.

When $\alpha\in(0, \frac{\ln(v_1^{(t)}/L)}{\ln(U/L)}]$ and $\alpha \in (\frac{\ln(v_{i-1}^{(t)}/L)}{\ln(U/L)},\frac{\ln(v_i^{(t)}/L)}{\ln(U/L)}]$, $i=2,\dots,n$, the reward function of this case is the same as that of Case (i). When $\alpha \in (\frac{\ln(v_n^{(t)}/L)}{\ln(U/L)},+\infty)$, we have $R_t(\cali_t,\alpha) = a_n^{(t)} - \frac{b_n^{(t)}}{\alpha}$ since $\Phi_\alpha(v_n^{(t)}) = \Phi_\alpha(U) = 1$.

Based on the closed-form expressions of the reward functions of \otp in Cases (i) and (ii), the following Lemma can be easily verified. 


\begin{lemma}
The reward function of \otp under the algorithm $\otpalg(\alpha)$ is Lipschitz-continuous in $\alpha\in(0,\infty)$. 
\end{lemma}

Combining the above lemma with the results of \cite{maillard2010online} yields a sublinear regret algorithm for \otp.  


\section{Online EV Admission Control (EAC)}
\label{app:ev}

In the online EV admission control (\EAC) problem, an operator of an EV charging station provides charging services for a sequence of EV customers. 
Due to the capacity limitation of the charging station, the operator needs to decide whether to admit each EV upon their arrivals to maximize the total values of all admitted EVs. 
In \EAC, we consider a time-slotted system with a time horizon $[T]:=\{1,\dots,T\}$. 
A set $[n] = \{1,\dots,n\}$ of EVs arrive for charging services. 
The charging request of EV $i\in[n]$ is characterized by $R_i = \{a_i,d_i,e_i;v_i\}$, where $a_i$ and $d_i$ denote the arrival and departure time slots, $e_i$ is the energy demand, and $v_i$ is the value of this charging request.
The charging station has a unit charging capacity $1$. 
Upon arrival of EV $i$, the operator receives its charging request $R_i$ and then makes two types of decisions (i) admission decision $x_i$, which determines whether to admit ($x_i = 1$) or reject ($x_i = 0$) this request, and (ii) schedule decision $\{y_{i,t}\}_{t \in T_i}$, where $y_{i,t}$ is the amount of energy delivered to EV $i$ at time slot $t$ and $T_i:=[a_i,\dots,d_i]$ is the feasible charging duration. 
The operator aims to maximize the total values of admitted EVs while ensuring that the capacity of the charging station is respected and all admitted EVs are delivered the required energy demand within its charging duration. The offline \EAC can be stated as

\begin{align}
{[\textrm{Offline EAC}]}& \quad \max\quad \sum_{i\in[n]}v_i x_i, \\
\label{eq:eac-energy}
{\rm s.t.}\quad& \sum_{t\in T_i} y_{i,t} \ge  e_i x_i, \quad \forall i\in[n],\\
\label{eq:eac-capacity}
&\sum_{i\in[n]} y_{i,t} \le 1, \quad \forall t\in[T],\\
{\rm vars.}\quad& x_i\in\{0,1\}, y_{i,t} \ge 0, \quad \forall i\in[n], t\in[T]. 
\end{align}
We assume that the energy demand of each EV is much less than the charging capacity of the station, i.e., $e_i \ll 1$, and the value density of each EV is upper and lower bounded, i.e., $L \le v_i/e_i \le U$. These two assumptions correspond to Assumptions \ref{assum:infitesimal} and \ref{assum:bounded-value-density} of \okp, respectively. 

The \EAC is a generalized version of \okp. Each EV corresponds to an item with value $v_i$ and weight $e_i$ in \okp. Instead of occupying the capacity of one knapsack in \okp, the item in \EAC can be divided into multiple knapsacks (each knapsack is indexed by $t\in T_i$). 
In what follows, we show a modified $\okpalg(\alpha)$ can solve the \EAC in online settings and ensures the same degradation factor as $\okpalg(\alpha)$.  

\subsection{Modified $\okpalg(\alpha)$}

\begin{algorithm}[!t]
	\caption{Modified $\okpalg(\alpha)$ for \EAC}
	\label{alg:eac}
	\begin{algorithmic}[1]
	    \State \textbf{Input:} threshold function $\Psi_\alpha(\cdot)$, initial utilization $z_{0,t} = 0, t\in[T]$;
		\While{EV $i$ arrives}
		\State determine a candidate charging schedule $\{\bar y_{i,t} \}_{t\in T_i}$ for EV $i$ by {\it water-filling schedule}
		\State determine $x_i^*$ by solving the following problem \begin{align}\label{p:eac}
        x_i^* = \argmax_{x_i\in\{0,1\}}\quad v_i x_i - \sum_{t\in T_i}\Psi_\alpha(z_{i-1,t})\bar y_{i,t},
        \end{align}
        \State update utilization $z_{i,t} = z_{i-1,t} + y^*_{i,t}$, where $y^*_{i,t} = x_i^* \bar y_{i,t}, t\in T_i$. 
		\EndWhile
	\end{algorithmic}
\end{algorithm}

We propose a modified $\okpalg(\alpha)$, depicted in Algorithm \ref{alg:eac}, to solve the \EAC. 
Let $z_{i,t}$ denote the utilization of the charging station at time $t$ after processing the previous $i$ EVs.
We use a threshold function $\Psi_\alpha(\cdot)$ to evaluate the marginal cost of charging at each time slot given current utilization of the station. For example, $\Psi_\alpha(z_{i-1,t})$ captures the marginal cost at time slot $t$ for EV $i$.

\paragraph{Water-filling Schedule.} Upon receiving the charging request $R_i$ from EV $i$, the algorithm first determines a candidate charging schedule based on water-filling. Particularly, the water-filling schedule $\{\bar y_{i,t}\}_{t\in T_i}$ is the solution of the following pseudo-cost minimization problem 
\begin{align}
    \min \quad&\sum_{t\in T_i} \int_{z_{i-1,t}}^{z_{i-1,t} + y_{i,t}}\Psi_\alpha(u)du,\\
    \label{eq:water-filling}
    {\rm s.t.}\quad& \sum_{t\in T_i} y_{i,t} \ge e_i,\\\nonumber
    {\rm vars.}\quad& y_{i,t} \ge 0, \forall t\in T_i.
\end{align}
In this schedule, the energy demand $e_i$ is continuously allocated to the time slot with the smallest marginal cost. This is why it is called a water-filling schedule.

\paragraph{Admission Control.} Note that the minimal cost of the water-filling schedule is $\sum_{t\in T_i}\int_{z_{i-1,t}}^{z_{i-1,t} +\bar y_{i,t}}\Psi_\alpha(u)du \approx  \sum_{t\in T_i} \Psi_\alpha(z_{i-1,t}) \bar y_{i,t}$ since $\bar y_{i,t} \le e_i \ll 1$. By using this cost as the estimated cost of charging EV $i$, the algorithm makes admission control by solving the pseudo-utility maximization problem \eqref{p:eac}. 

\paragraph{Updating Utilization.} Based on the admission decision, the final schedule decision is determined by $y_{i,t}^* = x_i^* \bar y_{i,t}$, i.e., using the candidate schedule if EV $i$ is admitted and setting a zero schedule otherwise. The algorithm finally updates the capacity utilization based on $y_{i,t}^*$.

\subsection{Degradation Factor Analysis of Modified $\okpalg(\alpha)$}

Our following analysis uses an online primal-dual approach. The proof of Theorem \ref{thm:DF_OKP} can be seen as a special case of the following proof. 

We relax the admission decisions from binary variables to continuous variables, i.e., $0\le x_i \le 1$.
The dual problem of the relaxed \EAC can be stated as
\begin{align*}
\min \quad& \sum_{t\in[T]} \mu_t + \sum_{i\in[n]} \eta_i\\
{\rm s.t.}\quad& \eta_i \ge v_i - \lambda_i e_i, \quad\forall i\in[n],\\
&\mu_t \ge \lambda_i, \quad\forall i\in[n],t\in[T],\\
{\rm vars.}\quad& \mu_t \ge 0, \eta_i \ge 0, \lambda_i \ge 0, \quad \forall i\in[n],t\in[T],
\end{align*}
where the dual variables $\{\lambda_i\}_{i\in[n]}$, $\{\mu_t\}_{t\in[T]}$, and $\{\eta_i\}_{i\in[n]}$ correspond to the constraints \eqref{eq:eac-capacity}, \eqref{eq:eac-energy}, and the relaxed constraint $x_i \le 1$, respectively.

Based on the admission decision $x_i^*$ and schedule decision $y_{i,t}^*$, produced by the modified $\okpalg(\alpha)$, we can construct a feasible dual variable by letting
\begin{align}
 \bar \mu_t &= \Psi_\alpha(z_{n,t}), \\ \bar\lambda_{i} &= \Psi_\alpha(z_{i-1,t} + y^*_{i,t}) \boldsymbol{1}_{x_i^* = 1, y^*_{i,t} > 0}, \\ \bar\eta_i &= (v_i - \bar\lambda_{i} e_i) \boldsymbol{1}_{x_i^* = 1},
\end{align}
where $\boldsymbol{1}_{\{\cdot\}}$ is the indicator function.
It is clear that the constructed dual variables $\bar \mu_t$ and $\bar\lambda_{i}$ are non-negative.
Note that $\bar\lambda_i$ is actually the optimal dual variable of the constraint \eqref{eq:water-filling} in the water-filling problem if EV $i$ is admitted and zero otherwise.
We have $\bar\lambda_i e_i = \sum_{t\in T_i} \Psi_\alpha(z_{i-1,t}+ y^*_{i,t})y^*_{i,t} \approx \sum_{t\in T_i} \Psi_\alpha(z_{i-1,t}) y^*_{i,t}$ since $y^*_{i,t} \ll 1$.
According to the decision rule of $x_i^*$ in Equation \eqref{p:eac}, $v_i - \bar\lambda_i e_i \ge 0$ when $x_i^* = 1$, and $v_i - \bar\lambda_i e_i < 0$ when $x_i^* = 0$. Thus, we have $\bar\eta_i \ge 0$ and the first dual constraint holds 
$$\bar\eta_i = (v_i - \bar\lambda_{i} e_i) \boldsymbol{1}_{x_i^* = 1} \ge (v_i - \bar\lambda_{i} e_i). $$
Since the threshold function is non-decreasing, we then have
\begin{align*}
  \bar \mu_t  = \Psi_\alpha(z_{n,t}) \ge \Psi_\alpha(z_{i-1,t} + y^*_{i,t}) \ge \bar\lambda_i.
\end{align*}
The last inequality holds since (i) if $y^*_{i,t} >0$, $\bar\lambda_i = \Psi_\alpha(z_{i-1,t} + y^*_{i,t})$, and (ii) if $y^*_{i,t} = 0$, $\bar\lambda_i < \Psi_\alpha(z_{i-1,t} + y^*_{i,t})$ based on the water-filling schedule.

The constructed feasible dual variables can build an upper bound for the offline optimum ${\sf OPT}$ of \EAC based on weak duality. Particularly, we have
\begin{small}
\begin{align*}
 {\sf OPT} &\le  \sum_{t\in[T]} \Psi_\alpha(z_{n,t}) + \sum_{i\in[n]} (v_i - \bar\lambda_{i} e_i)\boldsymbol{1}_{x_i^* = 1} \\
 &= \sum_{t\in[T]} \Psi_\alpha(z_{n,t}) +  \sum_{i\in[n]} v_i x_i^* - \sum_{i\in[n]}\sum_{t\in T_i} \Psi_\alpha(z_{i-1,t}+ y^*_{i,t})y^*_{i,t},  \\
 &\approx \sum_{t\in[T]} \left[\Psi_\alpha(z_{n,t}) - \sum_{i\in[n]}\int_{z_{i-1,t}}^{z_{i-1,t} + y^*_{i,t}}\Psi_\alpha(u)du \right]  + \sum_{i\in[n]} v_i x_i^*,\\
 &=\sum_{t\in[T]} \left[\Psi_\alpha(z_{n,t}) - \int_{0}^{z_{n,t}}\Psi_\alpha(u)du \right]  + \sum_{i\in[n]} v_i x_i^*,\\
 &\le \left[\frac{\texttt{DF}_\alpha}{T} - 1\right]\sum_{t\in[T]}  \int_{0}^{z_{n,t}} \Psi_\alpha(u)du + \sum_{i\in[n]} v_ix_i^*,\\
 &\le \left[\frac{\texttt{DF}_\alpha}{T} - 1\right] \sum_{i\in[n]} v_ix_i^* + \sum_{i\in[n]} v_ix_i^*,\\
 &=\frac{\texttt{DF}_\alpha}{T} \sum_{i\in[n]} v_ix_i^* = \frac{\texttt{DF}_\alpha}{T} {\sf ALG},
\end{align*}
where ${\sf ALG}$ is the total value achieved by the modified $\okpalg(\alpha)$. The first inequality comes from Equation \eqref{eq:DF-proof1} when $\alpha \in[1,\infty)$ and Equation \eqref{eq:DF-proof} when $\alpha \in (0,1)$. The second inequality is due to the fact that 
\begin{align*}
    \sum_{i\in[n]} v_ix_i^*  &\ge \sum_{i\in[n]}  \sum_{t\in T_i} \Psi_\alpha(z_{i-1,t}) y^*_{i,t} \\ &\approx 
    \sum_{i\in[n]} \sum_{t\in T_i} \int_{z_{i-1,t}}^{z_{i-1,t} + y^*_{i,t}}\Psi_\alpha(u)du \\
    &= \sum_{t\in[T]}  \int_{0}^{z_{n,t}} \Psi_\alpha(u)du.
\end{align*}
\end{small}
\section{Additional Experimental Details}
\label{app:exp}
The Caltech's ACN-data~\cite{LeeLiLow2019a} does not include the value of a charging request, an input parameter to the EV admission control problem. Hence, we use the following approach to estimate the value of each charging request. Specifically, we set the value for EV $i$ as $v_i = a (n_i + b e_i/(d_i-a_i))$, where $e_i$ is the energy demand, $d_i-a_i$ is the availability window, and $n_i$ is the estimated number of present users in the charging station at his arrival, and $a, b$ are constants. In our experiments, we set $a=12^{h}\times \texttt{unit\_price\_rate}$, and $b=2$ where \texttt{unit\_price\_rate} is an average of energy price during a day (approaximately \$0.06). To estimate $n_i$, we used the dataset to find the best Poisson distribution to describe the number of users present at the charging station as a function of time. For the online learning algorithm and in Figures~\ref{fig:regret_avg} and~\ref{fig:regret_val}, the learning parameter $\eta$ of the algorithm~\cite{maillard2010online} has been set to 0.9.

\end{document}